\documentclass[fleqn,11pt]{wlscirep}
\usepackage[utf8]{inputenc}
\usepackage[T1]{fontenc}
\usepackage{graphicx}
\usepackage{authblk}
\usepackage{xcolor}

\title{Autonomous Quantum Measurement with Agentic AI}
\author[1,2,$\dagger$]{Eunbi Rha}
\author[1,3,$\dagger$]{Abhijatmedhi Chotrattanapituk}
\author[1,4,5,$\dagger$]{Mouyang Cheng}
\author[3,$\dagger$]{Chung-tao Chou}
\author[6]{Shenyong Ye}
\author[3]{Ashley S. Dale}
\author[6]{Long Ju}
\author[3]{Luqiao Liu}
\author[1,2,4,*]{Mingda Li}
\affil[1]{Quantum Measurement Group, MIT, Cambridge, MA 02139, USA}
\affil[2]{Department of Nuclear Science and Engineering, MIT, Cambridge, MA 02139, USA}
\affil[3]{Department of Electrical Engineering and Computer Science, MIT, Cambridge, MA 02139, USA}
\affil[4]{Center for Computational Science \& Engineering, MIT, Cambridge, MA 02139, USA}
\affil[5]{Department of Materials Science and Engineering, MIT, Cambridge, MA 02139, USA}
\affil[6]{Department of Physics, MIT, Cambridge, MA 02139, USA}
\affil[$\dagger$]{These authors contributed equally.}
\affil[*]{e-mail: mingda@mit.edu}

\begin{abstract}
Quantum materials exhibit diverse electronic and functional phenomena, but their characterization requires extensive experimental expertise, iterative decision-making and accumulated laboratory knowledge that is difficult to document and transfer. 
Here we introduce Quantum Agent (QUANTA), an agentic artificial intelligence (AI) framework for autonomous quantum-materials characterization. 
Using a widely adopted Physical Property Measurement System (PPMS) as an experimental platform, QUANTA combines retrieval augmentation using scientific literature, experimental protocols and prior measurements with specialized agents for experimental planning, troubleshooting, and data interpretation. 
A separate deterministic control layer validates instrument actions. Guided by incoming data, QUANTA evaluates measurement quality, revises experimental plans and selects subsequent measurements. 
We demonstrate these capabilities through transport measurements of the topological Weyl semimetal TaP, as well as investigations of nonlinear responses in magnetic material Co$_2$MnGa. 
These results establish QUANTA as a generalizable framework for evidence-driven, safeguarded and closed-loop quantum-materials characterization.
\end{abstract}
\begin{document}

\flushbottom
\maketitle
\thispagestyle{fancy}

%%%%%%%%%%%%%%%%%%%%%%%%%%%%%%%%%%%%%%%%%%%%%%%%%%%%%%%%%%%%%%%%%%%%%%%%

\section*{Introduction}
Quantum materials exhibit diverse electronic and magnetic phenomena, ranging from unconventional superconductivity and quantum magnetism to topological electronic states
\cite{keimer2017physics,tokura2017emergent,basov2017properties,li2020thermal}. These phenomena arise from the interplay of electronic correlations, spin--orbit coupling and band topology, and can be investigated through variations in temperature and magnetic field, as well as carrier density and electric displacement field in suitably gated systems \cite{hasan2010colloquium,armitage2018weyl,burch2018magnetism,cao2018unconventional}.
Electrical transport is a central probe of quantum materials: longitudinal resistance tracks metallic, insulating and superconducting behavior and signatures of phase transitions \cite{imada1998metalinsulator};
Hall and magnetoresistance measurements probe carrier properties, magnetic responses and signatures associated with Berry curvature and band topology \cite{nagaosa2010anomalous}; and quantum oscillations constrain extremal Fermi-surface cross sections and cyclotron effective masses \cite{shoenberg1984magnetic}.

While transport measurements provide rich information about quantum materials, extracting reliable physical insights requires careful experimental design and interpretation. 
Even seemingly routine temperature or magnetic-field sweeps involve a series of sample-specific decisions. Researchers must select contact configurations and excitation levels, coordinate temperature and magnetic-field set points, assess sample equilibration and instrumental drift, and distinguish intrinsic material responses from measurement artifacts. 
For example, contact geometry can affect resistivity and Hall measurements \cite{vanderpauw1958method}, while alignment errors and inhomogeneous current flow can obscure intrinsic magnetotransport in high-mobility semimetals \cite{arnold2016negative,liang2018experimental,yang2019currentjetting}. Low-temperature, high-field measurements can also be time-consuming, and some fragile samples may degrade during repeated thermal cycling.
These constraints make experimental decisions consequential for both measurement efficiency and reproducibility. 
Recent advances in closed-loop laboratory automation offer a way to address some of these challenges, enabling experiments to adapt to incoming data \cite{hase2019nextgeneration,tom2024selfdriving,cheng2026artificial} and accelerating materials optimization, phase mapping and device tuning \cite{macleod2020selfdriving,kusne2020onthefly,moon2020machine}. 
Extending these capabilities to the measurement of quantum-materials requires integrating adaptive control with physical interpretation and troubleshooting across diverse samples and measurement protocols.

In this work, we introduce Quantum Agent (QUANTA), an agentic AI framework that integrates scientific reasoning with autonomous experimental control. 
QUANTA employs retrieval-augmented generation (RAG)\cite{lewis2020retrieval} to access scientific literature, instrument manuals, archived laboratory protocols and prior measurements, while specialized agents support experimental planning, data analysis and troubleshooting. 
A coordinating agent translates scientific objectives into structured, executable measurement plans and revises them based on experimental evidence. 
Implemented on a Physical Property Measurement System (PPMS) Dynacool from Quantum Design with external electrical instruments, QUANTA separates AI-based scientific decisions from deterministic instrument execution. 
Proposed actions are validated against instrument states and sample-specific constraints before execution by a persistent worker, with human intervention retained for physical operations. Measurement decisions, actions and results are recorded to ensure traceability. We demonstrate QUANTA across distinct experimental challenges in quantum materials. 
In the non-magnetic Weyl semimetal TaP, QUANTA qualifies measurement conditions, recovers from instrument faults and evaluates trade-offs between signal quality and acquisition time. 
In magnetic topological material Co$_2$MnGa, it adapts field-sweep protocols, investigates unexpected responses through reference measurements and examines nonlinear Hall signals without assuming their microscopic origin. 
These demonstrations show how agentic AI can advance quantum-materials characterization beyond predefined measurement routines, combining scientific reasoning, evidence-guided adaptive measurements and safeguarded experimental control to enable more systematic exploration of complex quantum phenomena.

\section*{Results}
\subsection*{Overview}
\begin{figure}[!htb]
    \centering
    \includegraphics[width=\linewidth]{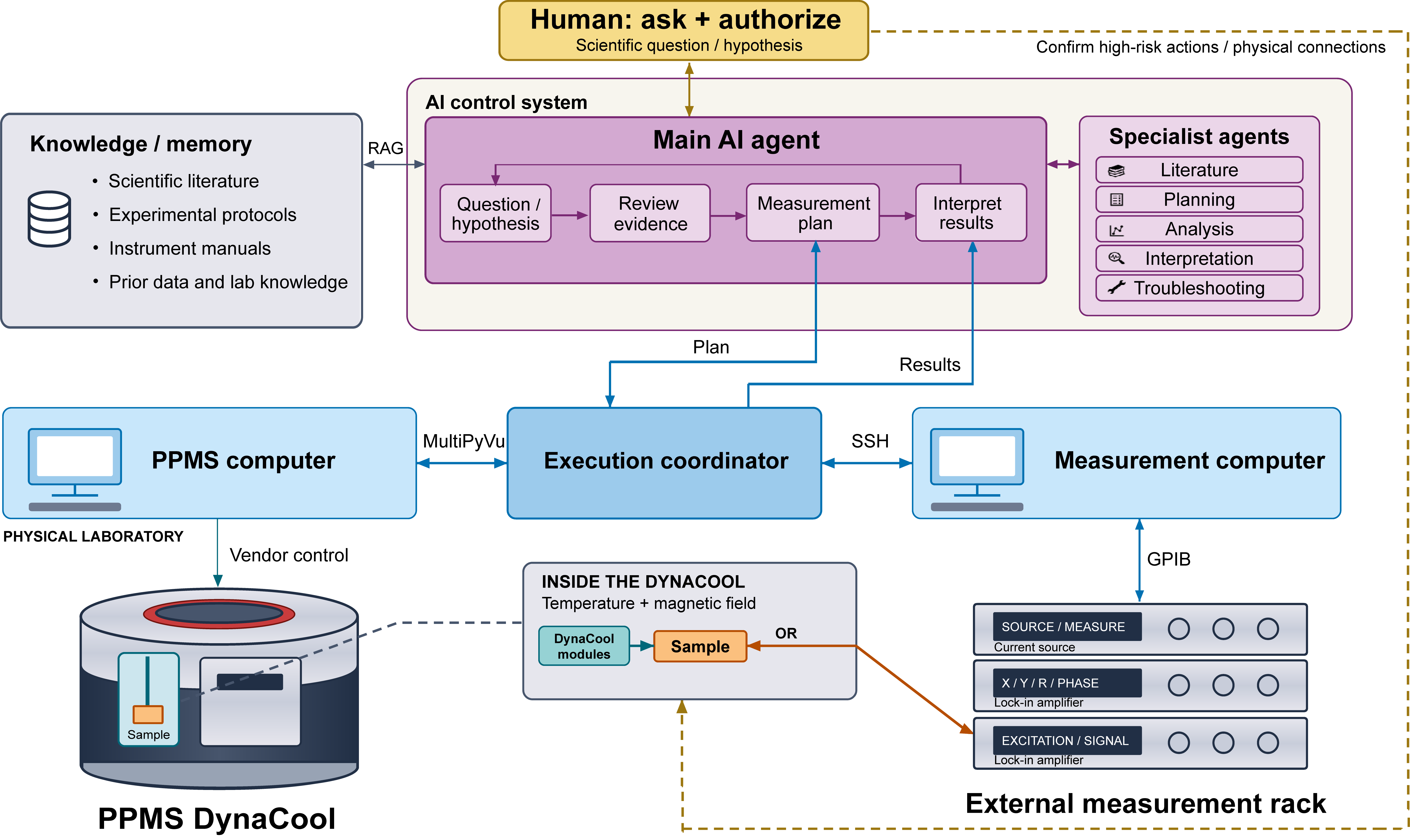}
    \caption{\textbf{Architecture of QUANTA for autonomous measurement of quantum materials.} A human researcher (top yellow block) supplies the scientific question and authorizes physical interventions. The main agent (purple block) iterates through evidence review, measurement planning and interpretation, drawing on literature, protocols, manuals and prior data in a shared knowledge store; specialist agents support these tasks. Plans pass to an execution coordinator (central blue block) for validation before a dedicated worker acts through the PPMS and measurement computers. Inside the DynaCool (bottom left), the PPMS sets temperature and field, while external current source and lock-in instruments provide excitation and electrical readout. Instrument data and faults return to the agent for the next decision. The dashed yellow arrow denote the researcher's physical oversight of the entire measurement process.}
    \label{fig:sdl_software_arch}
\end{figure}

QUANTA structures each measurement as a closed loop of question, evidence review, experimental planning, execution and interpretation (Fig.~\ref{fig:sdl_software_arch}). The researcher defines the objective, sample and experimental constraints. The main agent uses RAG to ground its reasoning in relevant literature, protocols, instrument manuals and prior runs, while specialist agents support planning, analysis and troubleshooting. The main agent retains the scientific decision of what to measure next to maximize information gain, resolve ambiguity and advance the experimental objective.

The main AI agent passes each proposed measurement to the execution coordinator, which serves as the interface between scientific reasoning and instrument control. The coordinator converts the proposed experiment into instrument-level actions and routes them to the appropriate control computer. Importantly, it also enforces deterministic safety constraints learned from instrument manuals, including hard limits on allowed set points, sweep ranges and instrument actions, so that the agent cannot directly issue commands outside a predefined safe operating envelope. Commands for measurement sequences, including temperature and magnetic-field control, are sent to the PPMS computer through MultiPyVu \cite{quantumdesign2017dynacool,quantumdesign2023multipyvu}, while transport measurements are coordinated through the measurement computer over SSH.
The PPMS computer communicates with the DynaCool system to execute temperature and field sweeps. In parallel, the measurement computer controls the current source and lock-in amplifiers through GPIB, a standard digital communication bus for computer-controlled laboratory instruments, coordinating excitation and voltage readout with the PPMS state (Supplementary Information Section 1). 
Subsequently, measurement results are passed back through the execution coordinator to the main AI agent, where they are interpreted together with the current experimental objective and retrieved knowledge to determine the next measurement. 
Throughout the whole process, human confirmation remains required for high-risk actions and physical interventions, including changes to wiring and sample connections.

This hierchical architecture of QUANTA converts potentially vague or tentative research prompts into a sequence of concrete, experimentally testable actions that are iteratively refined as new data arrive. By combining agentic scientific reasoning, automated experimental execution and deterministic safety control at the execution layer, QUANTA can autonomously navigate measurement workflows while keeping every action bounded, interpretable and traceable.

\subsection*{Autonomous calibration and selection of measurement parameters}
\begin{figure}[!htb]
    \centering
    \includegraphics[width=\linewidth]{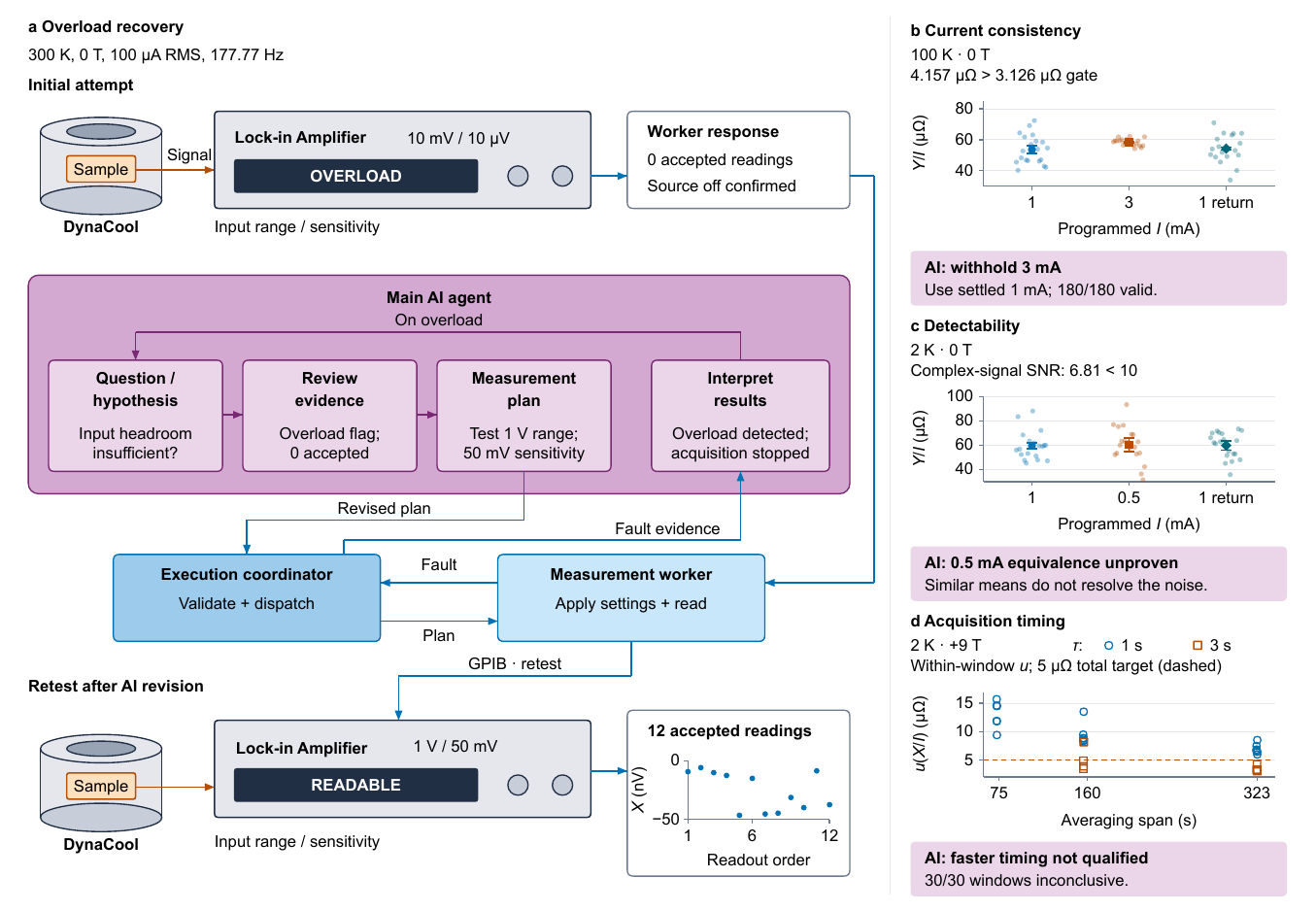}
    \caption{\textbf{Autonomous calibration of transport-measurement parameters by QUANTA.}
    \textbf{a.} Instrument-range recovery. A 100-$\mu$A root mean square (RMS) excitation at 300 K and 0 T produces a lock-in overload. QUANTA increases the lock-in input headroom, after which 12 readable measurements are recovered. The panel also shows the detailed closed-loop workflow, including fault reporting, agent reasoning, proposal of revised settings, coordinator validation and execution of the corrected measurement.
    \textbf{b.} Selection of excitation current for a consistent resistance measurement. At 100 K and 0 T, the resistance inferred at 3 mA differs from the settled 1-mA reference. QUANTA therefore rejects the 3-mA condition and retains 1 mA, for which all readings pass the validity checks.
    \textbf{c.} Selection of excitation current for high signal-to-noise ratio (SNR). At 2 K and 0 T, reducing the current to 0.5 mA gives a low complex-signal SNR of 6.81, below the acceptance threshold of 10. QUANTA therefore retains the higher-current condition.
    \textbf{d.} Selection of acquisition time duration. At 2 K and +9 T, 1-s and 3-s lock-in time constants are evaluated over increasing averaging spans against a 5-$\mu\Omega$ within-window uncertainty target. The faster acquisition does not satisfy the required precision, so QUANTA retains the longer measurement condition.}
    \label{fig:troubleshooting}
\end{figure}

QUANTA first establishes that a measurement is technically valid before using it for scientific interpretation. Fig.~\ref{fig:troubleshooting} illustrates four practical decisions that arise during transport measurements: whether the signal lies within the instrument range, which excitation current yields a consistent resistance, which current provides sufficient signal-to-noise ratio, and how long time the system must acquire data to reach the required precision.

The first test addresses a basic acquisition failure: whether the measured signal falls within the readable range of the lock-in amplifier. At 300 K and zero field, a transport measurement with a 100~$\mu$A root mean square (RMS) excitation produced an overload (Fig.~\ref{fig:troubleshooting}a), indicating that the signal exceeded the selected input configuration. 
The measurement worker therefore accepted no data, disabled the source and returned the fault state to the execution coordinator. This status was passed back to the main agent, which interpreted the event as an instrumental range problem rather than a material response and formulated a corrective action. 
It automatically proposed increasing the available input headroom, from a 10~mV range and 10~$\mu$V sensitivity to a 1~V range and 50~mV sensitivity. The proposal was then passed through the execution coordinator, which checked the requested settings against the permitted operating bounds before issuing the revised configuration to the measurement worker. 
The repeated measurement returned 12 readable values, which were reported back to the main agent and confirmed successful recovery of the acquisition process.

Having established a readable signal, QUANTA next determines an appropriate excitation current for extracting a reproducible resistance. Ideally, the measured voltage should scale consistently with the applied current over the selected operating range.
At $T=100$ K and zero magnetic field, the 3-mA run differed from the subsequent 1-mA return by 4.157~$\mu\Omega$, exceeding the 3.126~$\mu\Omega$ comparison threshold
calculated under the recorded uncertainty criterion 
(Fig.~\ref{fig:troubleshooting}b; Supplementary Information Section 2).
QUANTA therefore withheld the 3-mA condition and returned to the 1 mA setup, for which all 180 acquired readings passed the validity checks.
In this way, the agent automatically selects an excitation that preserves a consistent current--voltage response and therefore supports a reliable resistance estimate.

The third test asks a complementary question: how low can the excitation current be reduced while still retaining sufficient detectability? 
For a case at $T=$2 K and zero field, the mean response at 0.5 mA remained close to the 1 mA reference, but the complex lock-in signal became substantially noisier. 
This yields a signal-to-noise ratio of 6.81 below the acceptance threshold of 10, which is heuristically set by the QUANTA agent (Fig.~\ref{fig:troubleshooting}c). 
QUANTA therefore retained the 1 mA setting, which provided a more reliably detectable signal. Repeating the 1 mA condition after the low-current trial further tested whether the reference response remained stable during the comparison.

Finally, QUANTA determines how long each measurement must run to achieve the required precision. At $T=$2 K and $B=+9$ T, the agent compared 1 s and 3 s lock-in time constants over several averaging windows, using a 5~$\mu\Omega$ within-window uncertainty as the acceptance criterion (Fig.~\ref{fig:troubleshooting}d). 
The shorter setting did not reach the required precision in any of the 30 tested windows, so QUANTA retained the longer acquisition condition. This test allows the system to balance measurement speed against data quality, selecting an acquisition time that is sufficiently long for reliable statistics without unnecessarily prolonging the experiment.

\subsection*{Adaptive transport characterization of TaP}
\begin{figure}[!htb]
    \centering
    \includegraphics[width=\linewidth]{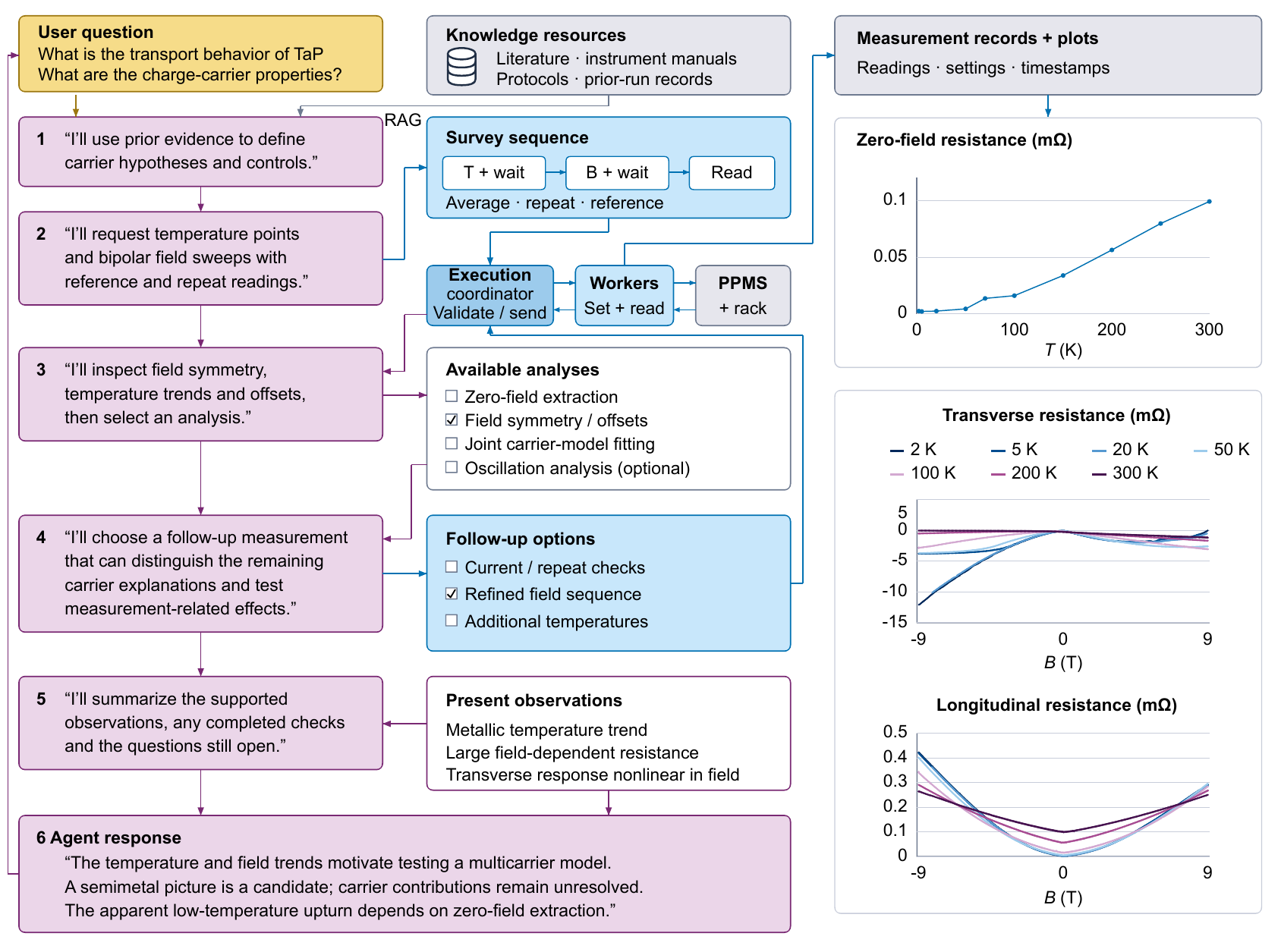}
    \caption{\textbf{Adaptive transport characterization of TaP.} The user question (yellow) and retrieved knowledge (grey) define candidate transport hypotheses and motivate an initial measurement survey (\textit{steps 1--2}). Temperature- and field-dependent resistance measurements are acquired through proposed sequences, and the resulting trends are used to select appropriate analyses (\textit{step 3}) and refine the next measurement sequence (\textit{step 4}). The example data show zero-field resistance versus temperature together with raw transverse and longitudinal field-dependent resistance in TaP, without transverse anti-symmetrization or longitudinal symmetrization. The transverse curves retain a substantial field-even contribution. Checked boxes denote agent-selected actions. Refined measurements are returned for execution, after which the agent summarizes supported observations and remaining uncertainties (\textit{step 5}) and formulates a final response that separates candidate interpretations from established findings (\textit{step 6}).}
    \label{fig:adaptive_transport}
\end{figure}

Fig.~\ref{fig:adaptive_transport} illustrates how QUANTA converts a scientific question into an adaptive measurement-and-analysis workflow, using our in-house transport measurements of TaP, a Weyl semimetal~\cite{arnold2016negative}, as an example. 
The user starts by asking about the transport behavior and charge-carrier properties of the sample. QUANTA then first combines this objective with retrieved literature, prior measurements and instrument knowledge to identify plausible carrier scenarios and the experimental controls needed to distinguish them (\textit{step 1}).
From this reasoning, the agent constructs an initial survey that samples resistance across temperature and performs bipolar magnetic-field sweeps at selected temperatures (\textit{step 2}). 
The measurements reveal three immediate features (Fig.\,\ref{fig:adaptive_transport}, right panels): a metallic temperature dependence of the zero-field resistance, a nonlinear transverse response and a strong longitudinal magnetoresistive response to magnetic field.

Rather than immediately assigning these signatures to a carrier model, QUANTA first determines which analyses are needed for reliable interpretation. The transverse and longitudinal field-dependent curves shown in Fig.\,\ref{fig:adaptive_transport} have not been anti-symmetrized or symmetrized, respectively. The transverse response contains a substantial field-even contribution, so its curvature alone does not establish multi-carrier transport; field-symmetry and offset analysis are needed before interpreting the intrinsic Hall response. The agent therefore selects field-symmetry and offset analysis from human-defined options (\textit{step 3}).
The resulting analysis then guides the next experiment. QUANTA selects a refined field sequence (\textit{step 4}) to better resolve the field-dependent response, while retaining current checks, repeat measurements and additional temperatures as alternative follow-ups if needed.
Using the refined data, the agent summarizes supported observations and remaining uncertainties (\textit{step 5}). The metallic temperature trend, strong field dependence and nonlinear transverse response are supported, whereas the detailed carrier contributions remain unresolved. The final response therefore identifies a semimetallic, potentially multicarrier picture as a candidate, while noting that the apparent low-temperature upturn depends on the zero-field extraction and that further measurements are needed to distinguish the remaining carrier scenarios (\textit{step 6}).

This example highlights QUANTA’s automated reasoning loop: each measurement aims to resolve a specific source of ambiguity, the returned data determine the appropriate analysis, and the resulting evidence guides the next experiment. 
In this way, QUANTA progressively narrows an open-ended scientific question into increasingly constrained hypotheses through iterative measurement, analysis and follow-up, while operating autonomously.

\subsection*{Investigating the nonlinear Hall response in Co$_2$MnGa}
\begin{figure}[!htb]
    \centering
    \includegraphics[width=\linewidth]{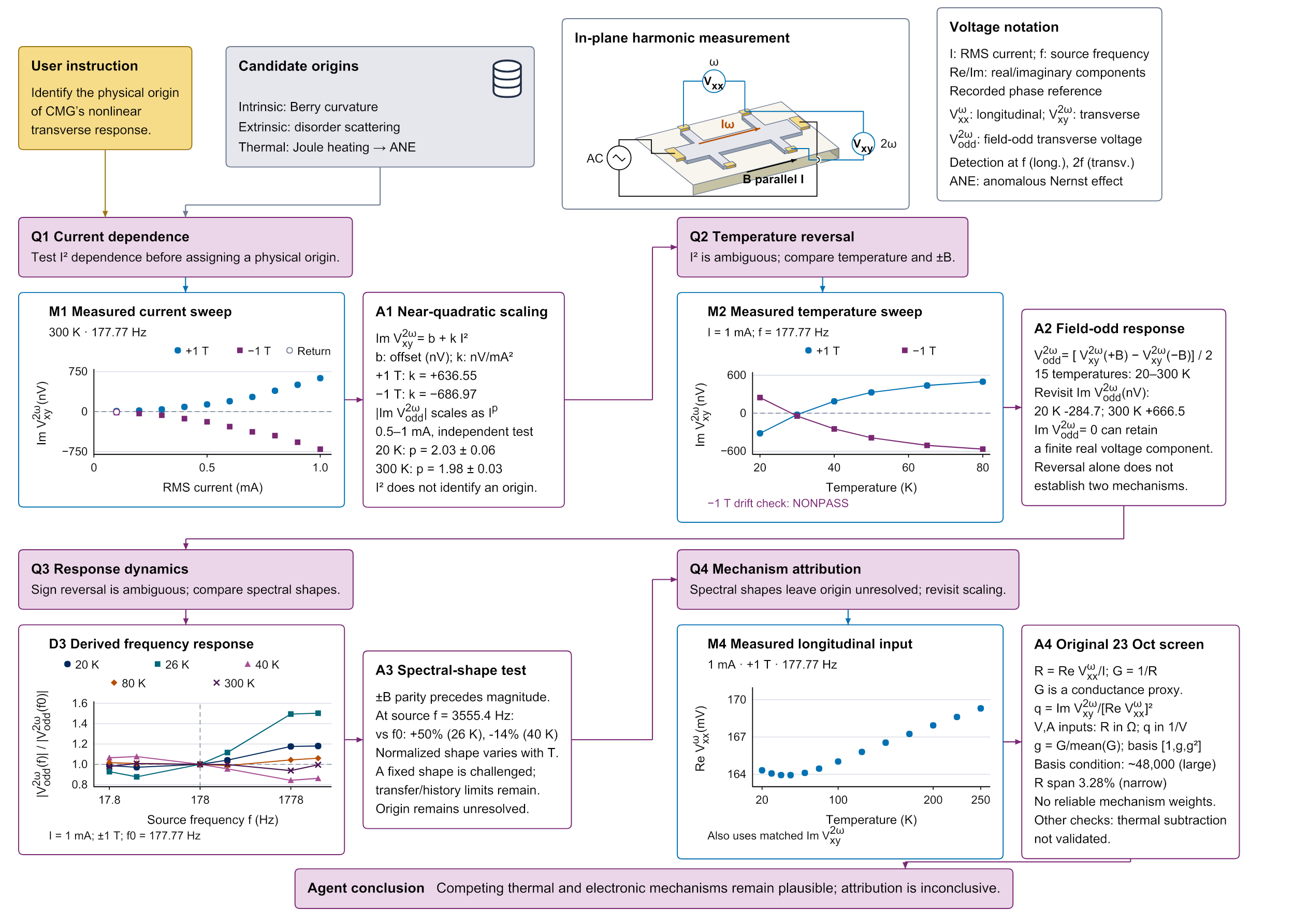}
    \caption{\textbf{Testing competing origins of the nonlinear transverse voltage in Co$_2$MnGa.} QUANTA progressively tests the mechanism assignment through four questions. 
    \textit{Q1}, current sweeps from 0.1--1~mA establish an approximately quadratic response, with fitted exponents close to 2, but this scaling is compatible with both electronic and Joule-heating-induced signals. 
    \textit{Q2}, temperature-dependent measurements reveal a sign reversal in the imaginary transverse component, although the full complex response remains finite and the reversal alone does not establish distinct mechanisms. 
    \textit{Q3}, frequency sweeps over 17.777--3555.4~Hz show temperature-dependent spectral shapes after normalization at 177.77~Hz, ruling out simple rescaling of a fixed response while remaining compatible with either multiple contributions or a single process with evolving dynamics. 
    \textit{Q4}, conductance-based scaling is strongly ill-conditioned because the longitudinal resistance spans only 3.28\% across the measured temperature range, preventing reliable separation of candidate contributions. 
    QUANTA therefore retains both electronic and thermoelectric origins as plausible and reports the microscopic attribution as inconclusive.}
    \label{fig:nonlinear_mapping}
\end{figure}

In Co$_2$MnGa, both nonlinear electronic transport and Joule-heating-induced thermoelectric effects can generate a transverse second-harmonic voltage, making their separation non-trivial (Fig.~\ref{fig:nonlinear_mapping}). 
This ambiguity is particularly relevant because Co$_2$MnGa exhibits a large anomalous Nernst response from the thermal effects \cite{guin2019anomalous}. We therefore ask QUANTA to determine the physical origin of the measured nonlinear transverse signal by progressively testing its current, temperature and frequency dependence. 
Measurements are performed in the nominal in-plane $I\parallel B$ geometry using the longitudinal first-harmonic voltage $V_{xx}^{\omega}$ and transverse second-harmonic voltage $V_{xy}^{2\omega}$.

As shown in Fig.~\ref{fig:nonlinear_mapping}, QUANTA first establishes the order of the nonlinear response (\textit{Q1}). Current sweeps over the measured range show near-quadratic scaling at both field polarities, and independent power-law fits at low and high temperatures give exponents close to two. 
The field-odd transverse component likewise follows approximately $I^2$. This confirms a robust second-order response but does not identify its origin, as both nonlinear electronic transport and Joule heating can produce the same current dependence.
QUANTA next tests whether the temperature dependence is consistent with a simple electrical-transport origin (\textit{Q2}). 
Although the longitudinal resistance changes only weakly with temperature, the imaginary component of the transverse second-harmonic voltage varies strongly and reverses sign.
Normalizing this response by the squared longitudinal voltage does not recover a simple conductance scaling with a fixed coefficient. 
However, the sign reversal of the imaginary component alone does not imply two distinct mechanisms, as a temperature-dependent phase shift can drive one component through zero while the total complex response remains finite. Together with stationarity issues in earlier negative-field measurements, QUANTA therefore treats the temperature dependence as evidence against the simplest transport-scaling picture, but not as sufficient for a unique microscopic assignment.

QUANTA next tests whether temperature changes only the response amplitude or also its dynamics (\textit{Q3}). Frequency sweeps across five temperatures show that, after normalization at a common reference frequency, the spectra do not collapse onto a single curve. Instead, their shapes change substantially with temperature. This rules out a simple picture in which temperature merely rescales a fixed frequency response, while remaining consistent with either multiple contributions or a single process with temperature-dependent dynamics.
QUANTA then asks whether these possible contributions can be separated from their temperature scaling (\textit{Q4}). To test this, the nonlinear transverse response is fitted using terms with different dependences on the longitudinal conductance, \(G=1/R_{xx}\), where \(R_{xx}\) is the measured longitudinal resistance. The conductance is normalized as \(g=G/\overline{G}\), where \(\overline{G}\) denotes its mean over the fitted temperature points, and the response is expanded in the basis \([1,g,g^2]\). However, \(R_{xx}\) changes by only a few percent over the measured temperature range, so \(g\) and \(g^2\) vary almost identically. The resulting fit is therefore strongly ill-conditioned: very different combinations of the three coefficients reproduce nearly the same temperature dependence. 
Consequently, QUANTA cannot reliably separate the candidate contributions from the present dataset and retains both electronic and thermoelectric origins as plausible, leaving the microscopic attribution inconclusive.

This experiment illustrates how QUANTA handles an underdetermined measurement without forcing a mechanistic assignment. 
It first establishes the nonlinear response, then successively tests current scaling, temperature dependence, spectral dynamics and model identifiability. Each step rules out progressively simpler explanations, while the remaining evidence is still insufficient to distinguish uniquely between electronic and thermal origins. 
QUANTA therefore carries this uncertainty into its final conclusion and identifies the additional controls needed for a more definitive assignment. Further examples of adaptive measurement planning and reference-assisted troubleshooting in Co$_2$MnGa are provided in Supplementary Information Section~3.

\section*{Discussion}
In this work, we developed QUANTA, an agentic framework that connects scientific reasoning with automated transport measurement in quantum materials. 
QUANTA translates open-ended scientific objectives into concrete actions, retrieves relevant literature and experimental knowledge, coordinates specialized agents for planning, analysis and troubleshooting, and selects subsequent measurements from the evidence acquired so far. 
Proposed actions are separated from instrument execution through a deterministic coordinator that enforces hard safety limits, while physical interventions remain under human authorization. 
This combination of agentic reasoning, automated experimentation and safeguarded control extends self-driving laboratory concepts~\cite{kusne2020onthefly,abolhasani2023rise,tom2024selfdriving} toward open-ended scientific reasoning (Supplementary Information Section 4). 
Across the topological materials TaP and Co$_2$MnGa, QUANTA performs adaptive transport characterization and troubleshooting, and proposes follow-up measurements to distinguish competing mechanisms underlying the observed transport behavior.

Several challenges nevertheless remain. 
First, QUANTA is deliberately conservative when the available measurements do not uniquely support an interpretation. This limits unsupported claims, but can also make it reluctant to formulate bold hypotheses from incomplete evidence. 
Future systems will need to more clearly distinguish hypotheses proposed for further testing from conclusions already supported by the available evidence.
Automation also currently ends at operations that can be controlled electronically: sample mounting, rewiring and contact changes still require human intervention, and extending the loop to these tasks will require robotic handling and laboratory automation~\cite{burger2020mobile,szymanski2023autonomous,tom2024selfdriving}. More generally, even with these advances, human researchers must remain responsible for the scientific interpretation and claims drawn from the experiment.

The same architecture of QUANTA could extend far beyond PPMS-based transport. Within quantum materials, agentic control could be coupled to dilution refrigerators, electrostatic gates, microwave electronics and scanning probes for strongly correlated systems and low-dimensional materials. 
More broadly, the principle of selecting the next measurement to maximally reduce scientific uncertainty is independent of experimental modality. Adaptive data acquisition has already been demonstrated in X-ray diffraction~\cite{szymanski2023adaptively}, synchrotron scattering~\cite{kusne2020onthefly,yager2023autonomous,cao2026uncertainty} and neutron diffraction~\cite{mcdannald2022fly}, with recent work extending LLM-based agents to synchrotron operation~\cite{chen2026agentic}. 
Similar strategies could guide optical spectroscopy, X-ray absorption spectroscopy, inelastic neutron scattering, X-ray photon correlation spectroscopy and other multimodal probes. 
Coupled to stronger theoretical and computational agents, this could enable a closed loop in which hypotheses are proposed, tested experimentally and refined against theory and simulation, opening a path toward autonomous scientific discovery in condensed-matter physics.

\section*{Methods}
\subsection*{Instrument control and run record}
QUANTA uses a coordinator on a controlling computer and a separate execution path for the apparatus. The AI model and its roles in the demonstrated campaigns are described in Supplementary Information Section 7. Each proposal is serialized as a structured request specifying the sample identifier, target temperature and field, stabilization criteria, excitation and lock-in configuration, acquisition window and abort conditions. The execution coordinator checks types, numerical bounds, requested command order and current instrument state before dispatch. It records both accepted and rejected requests with their stated rationale. A persistent worker is the sole owner of the live hardware sessions; it records raw readings, instrument settings, status flags, timestamps and worker events in a run database. Temperature and magnetic-field commands reach the DynaCool through the PPMS computer and vendor-control software. A Keithley 6221 supplies AC current; two SR860 lock-in amplifiers read voltage channels through a measurement computer. The researcher confirms measurement specific sample mounting, wiring and any intervention that cannot be verified through the instrument interface. Instrument control used MultiPyVu 2.1.4 for the DynaCool and PyVISA 1.16.2 with PyMeasure 0.16.0 over GPIB for the current source and lock-in amplifiers, with commands constrained by campaign-specific temperature, magnetic-field, ramp-rate, current and compliance limits. Both SR860s used differential, AC-coupled voltage inputs and an external reference from the Keithley 6221 phase marker; voltage-to-resistance conversion used the programmed RMS current amplitude.

For each run, the stored plan is linked to the resulting data and any agent interpretation. Set-point requests and measured temperature and field are retained separately, so an unsettled cryostat is not represented as being at the requested condition. The worker records the source-off confirmation after a fault and requires a newly validated request before a retest. The scientific record includes prompts or decision summaries, retrieval identifiers, analysis-code version and operator overrides alongside the raw traces. The committed records also support a read-only dashboard
for reviewing measurement progress and cached instrument
status (Supplementary Information Section 5). The archived laboratory-sequence retrieval corpus,
its procedural coverage and potential data leakage
are detailed in Supplementary Information Section 6.

\subsection*{Quality gates and transport analysis}
For every acquisition window, the worker saves the lock-in \(X\), \(Y\), magnitude, phase and status flags before a reading is accepted. An overload, communication fault or source inconsistency invalidates the affected window; the current source is turned off when required by the fault policy. Current consistency is assessed by comparing inferred resistance at the candidate excitation with a settled reference and a return measurement. Detectability is evaluated from the complex response rather than magnitude alone, and precision from the within-window uncertainty in \(X/I\). The Fig.~\ref{fig:troubleshooting} gates are 3.126 \(\mu\Omega\) for current consistency, a complex-signal signal-to-noise ratio of 10, and 5 \(\mu\Omega\) for within-window uncertainty (see more details in Supplementary Information Section 2). 

The field-dependent curves displayed in Fig.\,\ref{fig:adaptive_transport} retain both field-even and field-odd contributions: no longitudinal symmetrization or transverse anti-symmetrization was applied. Field-symmetry and offset analysis are identified as follow-up analyses rather than completed processing of these curves. In particular, the longitudinal trace is symmetrized as \(\rho_{xx}^{\mathrm{even}}(B)=[\rho_{xx}(B)+\rho_{xx}(-B)]/2\) and the Hall trace anti-symmetrized as \(\rho_{xy}^{\mathrm{odd}}(B)=[\rho_{xy}(B)-\rho_{xy}(-B)]/2\), after aligning the two field grids. Multi-carrier fits use a specified field and temperature interval and report covariance or resampling-based uncertainty; fits are not used to establish a carrier count when parameters are non-identifiable. 
For quantum oscillations, a smooth background is removed from the low-temperature field trace, the residual is resampled uniformly in \(1/B\), and the spectrum is compared across background choices and inverse-field windows. Effective masses are obtained from temperature-dependent oscillation amplitudes only when a peak can be tracked reproducibly.

Harmonic measurements retain complex \(V_{xx}\) and \(V_{xy}\) at \(\omega\) and \(2\omega\), together with drive current, phase reference, frequency and lock-in time constant. Quadratic-current tests are fitted over the drive interval that passed the ordinary transport quality gates; the intercept and fit residuals are reported. Lead interchange, field reversal, frequency and angular tests are analysed separately because electrical nonlinearity, thermal gradients and linear-channel leakage need not transform in the same way.

\subsection*{Sample fabrication}
Resistivity conversion requires measured sample dimensions and the current and voltage-contact geometry for each specimen. The crystallographic current and field directions, sample provenance and contact preparation are recorded with the corresponding run. The TaP sample was prepared according to Ref.\cite{han2020quantized}. The 30nm Co$_2$MnGa sample was deposited by magnetron sputtering on MgO(001) and capped with 10 nm of sputtered MgO, followed by hall bar patterning with a lithography process. 

\subsection*{Evaluation of autonomous decisions}
A quantitative comparison defines the same starting information, sample access and stopping criterion for agent-guided and human-operated campaigns. Completion is counted only when the requested feature is supported by qualified data and the prespecified analysis. Additional outcomes are elapsed instrument time, number of field and temperature transitions, rejected measurement windows, operator interventions and agreement on extracted transition temperatures, oscillation frequencies or map boundaries. Fault recovery is scored separately from scientific interpretation, since a readable retest does not prove that the original fault was diagnosed. Further details of data provenance, interpretive limitations
and agent computational usage are provided in
Supplementary Information Section 7.

\section*{Funding}
ER acknowledges support from the U.S. Department of Energy (DOE), Office of Science, Basic Energy Sciences (BES), under Award No. DE-SC0020148. AC acknowledges support from the U.S. National Science Foundation (NSF) Convergence Accelerator Program under Award No. ITE-2345084. CC and LL acknowledge support from National Science Foundation (NSF) under award no. DMR2522384.

\section*{Acknowledgements}
ER also acknowledges support from the MathWorks Fellowship. MC acknowledges support from the MIT Energy Initiative (MITEI) Fellowship. ML acknowledges the support from R Wachnik. 

\section*{Competing interests}
The authors declare no competing interests.

\bibliography{refs}

\end{document}

\maketitle

\tableofcontents

\section{Experimental Platform and Instrumentation}
\label{sec:experimental_platform}

\subsection{Hardware}
\label{sec:hardware}

\begin{figure}[htbp]
    \centering
    \includegraphics[width=0.9\linewidth]
        {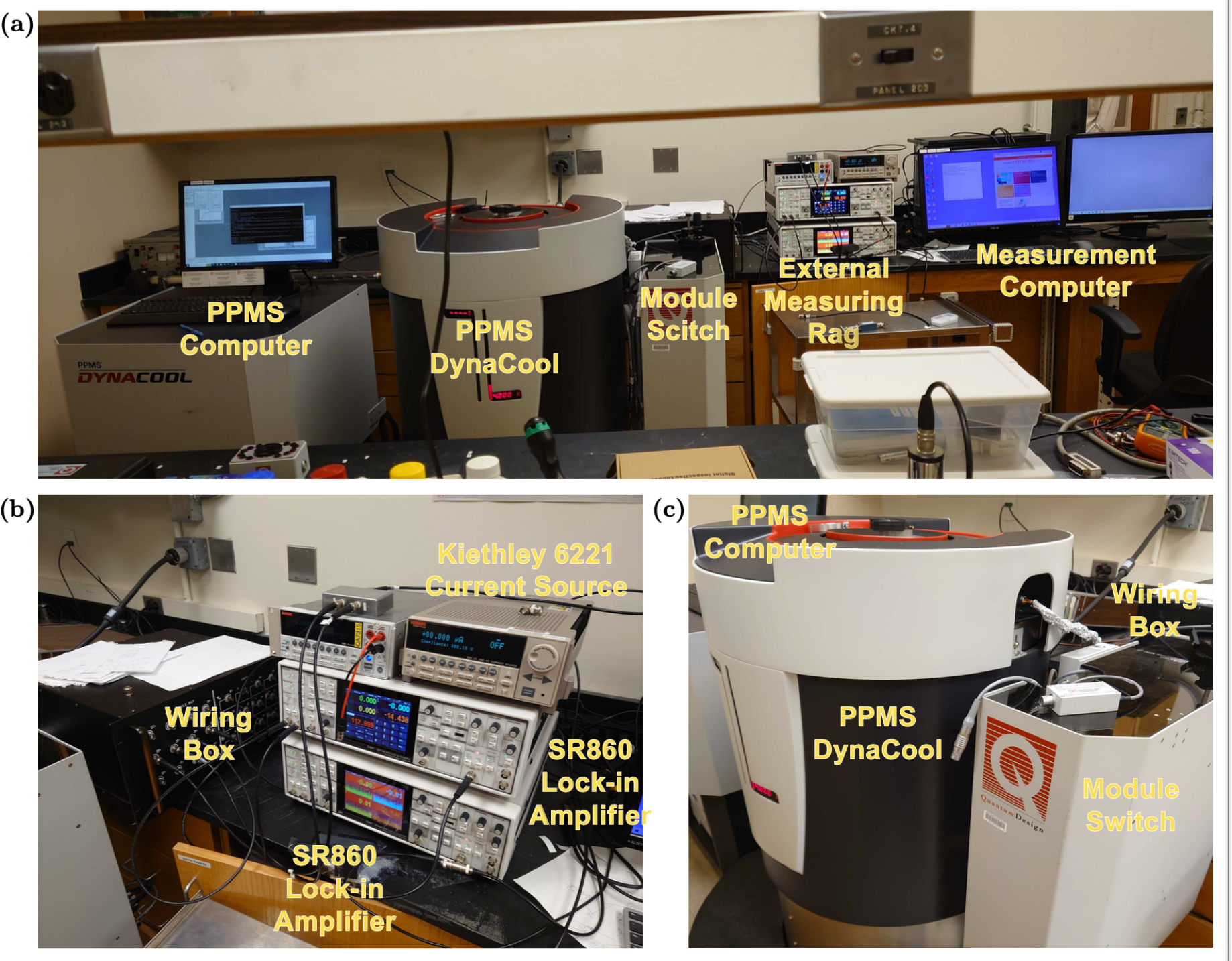}
    \caption{
        \textbf{Physical configuration of the
        agentic PPMS experimental platform.}
        \textbf{(a)} Laboratory overview showing
        the PPMS control computer, DynaCool,
        external measurement rack, module switch,
        and measurement computer.
        \textbf{(b)} Close-up of the external
        measurement electronics, including the
        Keithley 6221 current source, two SR860
        lock-in amplifiers, and sample wiring box.
        \textbf{(c)} DynaCool sample-interface
        region and adjacent connection hardware.
        The photographs identify the physical
        components corresponding to the control
        and measurement paths in main-text Fig.~1;
        individual sample-contact assignments
        depend on the experimental wiring.
    }
    \label{fig:ppms_hardware}
\end{figure}

The experimental platform combines a Quantum Design
PPMS DynaCool with external electrical transport
instruments and a separate measurement computer
(Supplementary Fig.~\ref{fig:ppms_hardware}).
The DynaCool provides the controlled temperature
and magnetic field environment. External electronics
supply current excitation and measure the sample
voltages. This arrangement separates environmental
control from electrical acquisition while allowing
the measurement software to coordinate both.

The system uses three computers with distinct
responsibilities. The researcher's computer hosts
the QUANTA coordinator and provides the interface for
defining experimental objectives, reviewing plans,
and inspecting results. The measurement computer
hosts the MCP control gateway and a persistent
experiment worker. A dedicated PPMS computer runs
MultiVu and the MultiPyVu server. The coordinator
accesses the control gateway over SSH, while a
MultiPyVu client on the measurement computer
communicates with the PPMS server over TCP.

The external transport electronics comprise a
Keithley 6221 current source and two Stanford
Research Systems SR860 lock-in amplifiers.
The measurement computer controls these
instruments through NI-VISA/GPIB.
The 6221 supplies AC current excitation, and
the two SR860s detect separate sample-voltage
channels, enabling longitudinal and transverse
measurements when the corresponding contacts
are connected. A wiring box provides access to
the sample connections. The assignment of each
voltage channel depends on the wiring used for
the experiment.

Instrument communication and phase synchronization
use separate connections. GPIB carries configuration
commands and measurement readbacks. A physical
reference signal from the 6221 phase-marker output
provides synchronization for the lock-in amplifiers.
Before acquisition, the setup checks include
instrument identities, source amplitude and
compliance, lock-in reference lock, input overload,
and the correspondence between voltage inputs
and sample contacts. Current amplitudes must
explicitly distinguish peak and RMS conventions.

The persistent worker coordinates environmental
settling, electrical acquisition, and data recording.
It owns the hardware connections independently
of the coordinator's SSH session, allowing a
submitted run to continue through a coordinator
disconnection according to its configured execution
policy. The run record stores plans, execution
status, and measurements. Analysis routines and
the read-only dashboard use these records to
support inspection and subsequent experimental
decisions.

\subsection{Software and data architecture}
\label{sec:software_architecture}

\begin{figure}[p]
    \centering
    \includegraphics[
        width=\linewidth,
        height=0.82\textheight,
        keepaspectratio
    ]{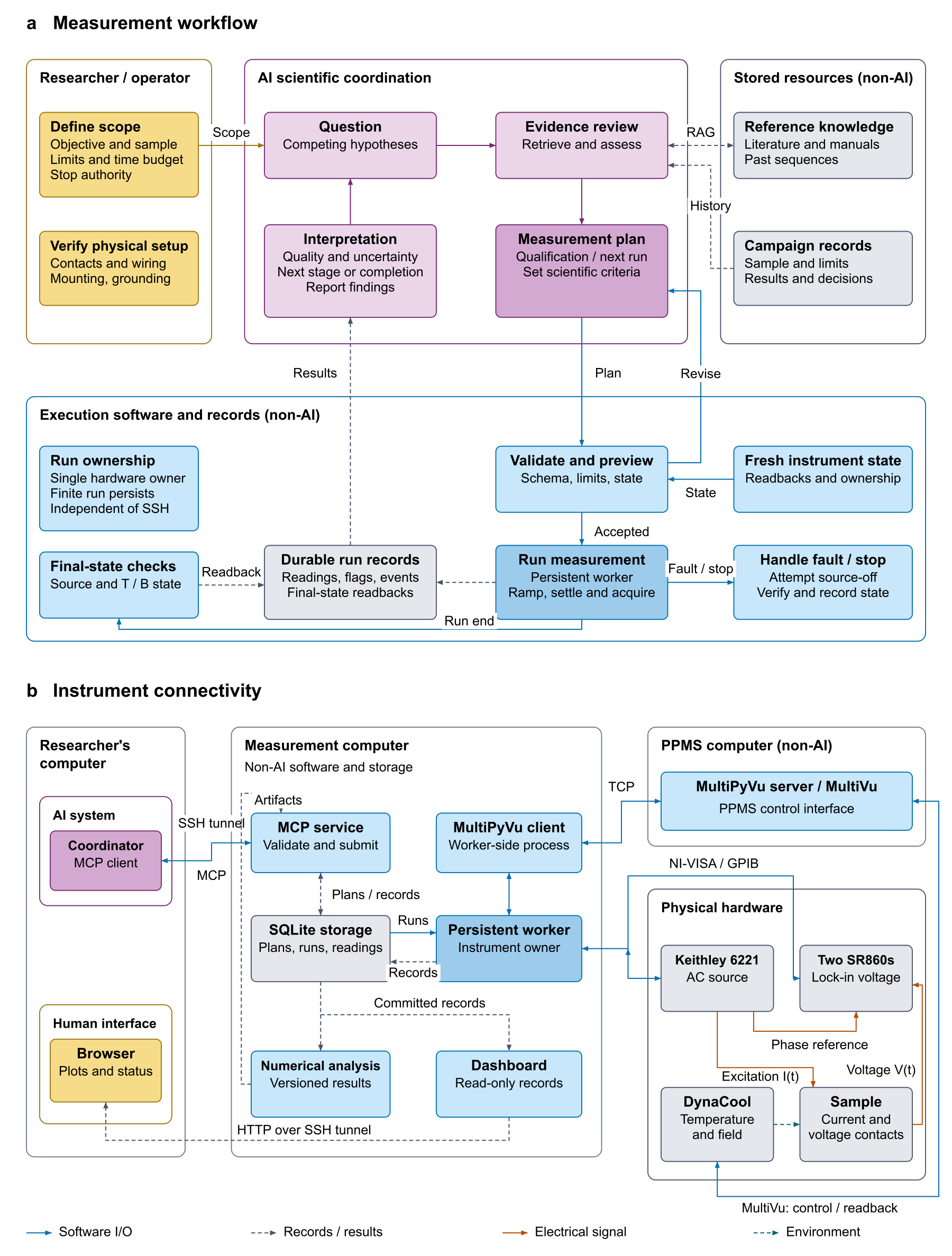}
    \caption{
    \textbf{Scientific workflow and instrument connectivity of QUANTA.}
    \textbf{(a)} Agent-guided measurement workflow, distinguishing
    researcher-defined objectives and physical verification, AI-based
    scientific coordination, stored knowledge and campaign records,
    and deterministic execution software. The AI coordinator retrieves
    evidence, formulates measurement plans, and interprets results,
    while software validation and a persistent worker enforce the
    authorized operating envelope and record experimental outcomes.
    \textbf{(b)} Connectivity among the researcher's computer,
    measurement computer, PPMS control computer, and physical
    instruments. The AI coordinator communicates with the MCP
    service over SSH, while the persistent worker interfaces with
    the PPMS through MultiPyVu and with the external electrical
    instruments through NI-VISA/GPIB. SQLite storage supports
    persistent execution records, numerical analysis, and a
    read-only dashboard. Arrows distinguish software communication,
    stored records, electrical signals, and the sample environment.
    }
    \label{fig:si_workflow_connectivity}
\end{figure}

QUANTA separates scientific coordination from instrument execution.
The overall measurement workflow is illustrated in Supplementary
Fig.~\ref{fig:si_workflow_connectivity}a. The researcher defines the
scientific objective, operating limits and stop authority, and verifies
the physical setup. Within these boundaries, the AI coordinator
retrieves evidence, generates measurement plans, interprets results
and selects subsequent measurements, while deterministic software
handles validation, instrument execution and data recording.
Supplementary Fig.~\ref{fig:si_workflow_connectivity}b summarizes
the connectivity among the researcher, measurement and PPMS
computers and the physical instruments. The detailed software
architecture, including execution ownership, safety enforcement
and durable records, is shown in Supplementary
Fig.~\ref{fig:si_system_architecture}.

The coordinator does not hold the instrument sessions. Each
executable plan is a structured object containing the sample
identifier, ordered temperature and field points, excitation,
channel settings, settling and repetition parameters, abort
conditions and an idempotency key. The application layer
validates the object against the deployed schema and numerical
limits before placing it in a SQLite-backed queue. A persistent
worker is the sole owner of the live instrument connections.
This prevents an interrupted SSH or model session from leaving
two independent clients competing for the same instruments.

The worker commits five classes of durable record: immutable
plans, run state, raw readings, timestamped events and cached
instrument state. Plan and configuration hashes link a run to
the exact action envelope under which it was validated.
Requested set points are stored separately from measured
temperature and field. Each electrical reading retains the
lock-in $X$ and $Y$ components, status flags, excitation
settings and environmental readbacks; rejected or faulted
readings therefore remain available for audit rather than
disappearing from the scientific record. The read-only
dashboard and numerical analysis processes consume committed
records and cannot dispatch instrument actions.

The worker may continue a finite, already dispatched run if
the coordinator disconnects. Advancement to a new scientific
stage still requires the coordinator or a separately validated
finite controller. The current implementation is therefore
persistent at the run level, but is not a generic unattended
campaign scheduler.

\clearpage
\begin{figure}[p]
    \centering
    \includegraphics[width=1.0\linewidth,keepaspectratio]
        {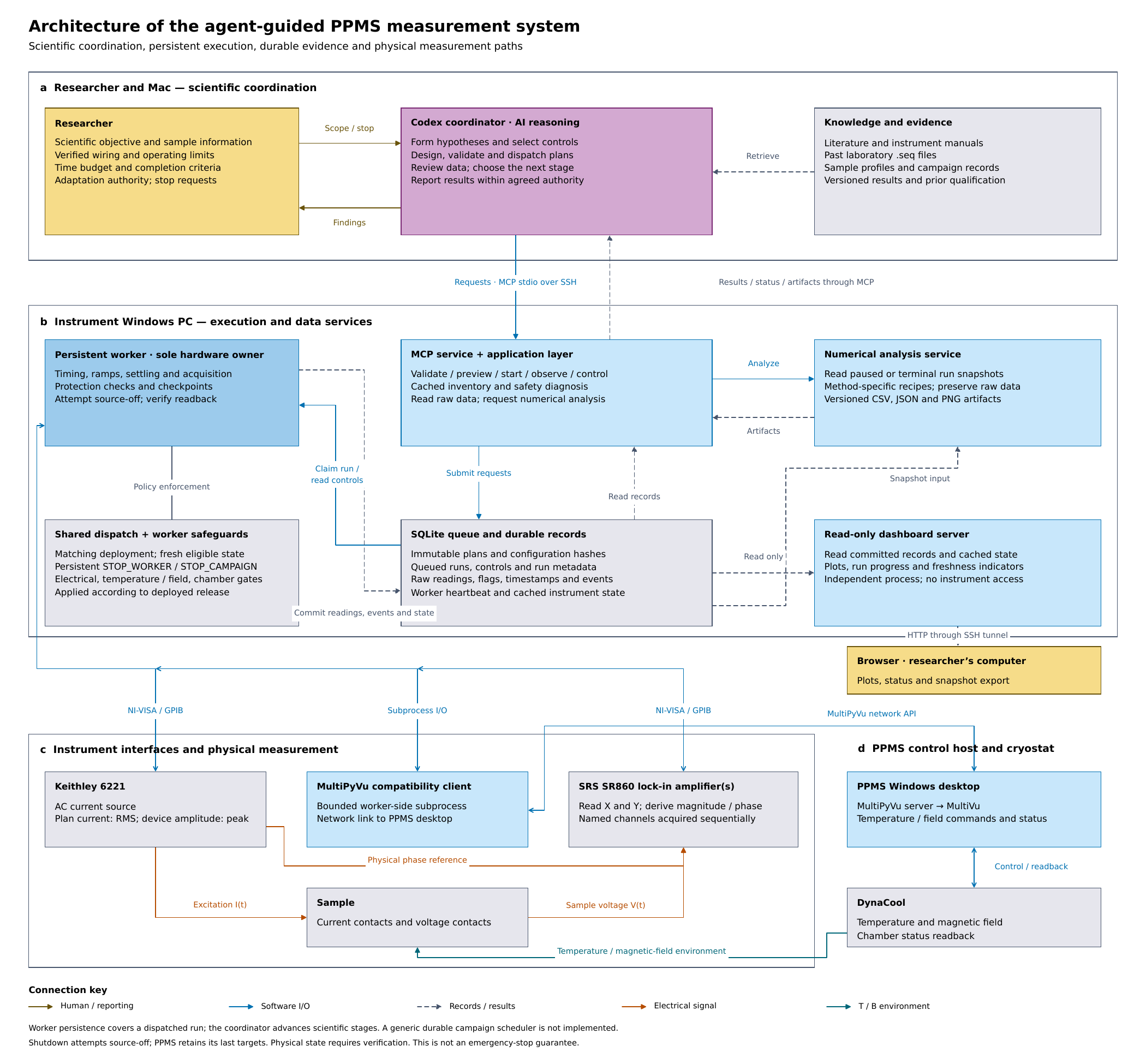}
    \caption{
    \textbf{Detailed software architecture, execution safeguards
    and data provenance.}
    The implementation-level architecture underlying the workflow
    in Supplementary Fig.~\ref{fig:si_workflow_connectivity}.
    \textbf{(a)} The researcher defines the objective, operating
    envelope and stop authority, while the AI coordinator retrieves
    evidence and selects measurements.
    \textbf{(b)} Structured requests pass through application-level
    validation and a durable queue before execution by the
    persistent worker, which maintains exclusive instrument
    ownership. Deployment checks, stop controls, instrument-state
    safeguards and immutable run records support bounded execution
    and auditability.
    \textbf{(c)} NI-VISA/GPIB carries instrument commands and
    readbacks, separately from excitation, voltage and
    phase-reference signal paths.
    \textbf{(d)} MultiPyVu connects the worker to MultiVu and
    the DynaCool. Run-level persistence does not constitute an
    unattended campaign scheduler; source-off attempts do not
    zero the magnet or cancel retained PPMS targets.
    }
    \label{fig:si_system_architecture}
\end{figure}
\clearpage

\section{Agent Instructions and Auditable Decision Records}
\label{sec:agent_records}

\subsection{Instruction hierarchy}

The operating instructions are divided into layers so that a general language
model request cannot silently become an instrument command. The principal layers
and the information passed between them are summarized in Supplementary
Table~\ref{tab:instruction_hierarchy}. The researcher defines the scientific
target, the sample-specific bounds and the permitted degree of adaptation. The
repository entry instructions require the coordinator to identify the exact
deployment, worker owner, state directory, instrument identities and wiring
record before any action. The operating guide then defines evidence retrieval,
minimum sufficient qualification, plan validation, monitoring, fault handling
and reporting. Campaign records preserve the scope and scientific state, whereas
machine-readable plan and configuration files contain the executable values.

\begin{table}[htbp]
\centering
\caption{Instruction and record hierarchy used by QUANTA.}
\label{tab:instruction_hierarchy}
\small
\begin{tabularx}{\linewidth}{@{}P{0.25\linewidth}Y@{}}
\toprule
Layer & Required content and role \\
\midrule
Researcher intake & Objective, sample identity, physical wiring evidence,
allowed temperature, field and excitation range, time budget, completion
criterion, excluded physical actions and stop authority. Establishes campaign
scope but does not encode an executable command \\
Repository entry instructions & Exact deployment, worker owner, configuration,
state directory and requirement to read the operating guide. Prevents reuse of
a stale endpoint or a different sample's profile \\
Operating guide & Manual and literature retrieval, sequence search, qualification,
validation, monitoring, troubleshooting and shutdown policy. Defines how a
scientific objective is converted into bounded stages \\
Campaign record and checklist & Hypotheses, controls, authorization, limits,
accepted and rejected evidence, pending route and final-state requirement.
Maintains the evolving scientific state and handoff record \\
Plan and deployment configuration & Ordered set points, rates, electrical
settings, repetitions, stop conditions, numerical ceilings and configuration
hash. Supplies the only representation accepted by the validator and worker \\
SQLite run record & Plans, runs, readings, events, cached state, timestamps,
quality flags and shutdown result. Connects a scientific decision to the exact
action and returned evidence \\
\bottomrule
\end{tabularx}
\end{table}

\subsection{TaP qualification metrics and acceptance criteria}
\label{sec:tap_qualification}

TaP qualification separated acquisition validity, signal detectability,
stationarity and reproducibility. Numerical acceptance criteria were
operational choices recorded for the campaign, rather than intrinsic
material constants or damage thresholds. In particular, the factors
of three, the 1\% comparison floor, the detectability cutoff of 10 and
the later $5\,\mu\Omega$ precision target were selected criteria;
the measured means, uncertainties and resulting numerical comparison
thresholds were calculated from the acquired data. The early
qualification checks used descriptive block statistics, whereas the
later timing study used a separately specified correlation-aware
protocol.

\paragraph{Signal normalization and block uncertainty.}
For each accepted reading, the in-phase and quadrature voltages were
normalized by the programmed RMS excitation current:
\begin{equation}
r_{X,i}=\frac{X_i}{I_{\mathrm{RMS}}}, \qquad
r_{Y,i}=\frac{Y_i}{I_{\mathrm{RMS}}}.
\end{equation}
These quantities have units of resistance, but normalization by
programmed current does not constitute an independent calibration of
the actual sample current. The phase convention was retained between
comparisons, and $Y/I$ was not interpreted as a Hall resistance.

For the early 20-reading qualification records, consecutive readings
were divided into four nonoverlapping blocks of five. For component
$a\in\{X,Y\}$, let $b_{a,j}$ denote the mean of block $j$. The run mean
and descriptive standard uncertainty were
\begin{equation}
\bar r_a=\frac{1}{4}\sum_{j=1}^{4}b_{a,j},
\qquad
u_a=\frac{s(b_{a,1},\ldots,b_{a,4})}{\sqrt{4}},
\end{equation}
where $s$ is the sample standard deviation with three degrees of
freedom. Blocking reduced reliance on treating individual lock-in
readings as independent, but four blocks did not establish that
residual correlation or drift was negligible.

\paragraph{Current-comparison threshold.}
For two runs $A$ and $B$, the componentwise difference and its
descriptive combined uncertainty were
\begin{equation}
\Delta_a=\bar r_{a,B}-\bar r_{a,A},
\qquad
u_{\Delta,a}=\sqrt{u_{a,A}^{2}+u_{a,B}^{2}}.
\end{equation}
The recorded comparison rule used the larger of
$3u_{\Delta,a}$ and a 1\% signal-scale floor. The multiplier and
percentage floor were operational screening choices; this rule was
not a formal demonstration of 1\% equivalence.

For the 100-K, zero-field 1--3--1 mA bracket, the quoted comparison
was between the 3-mA run and the subsequent 1-mA return run.
For $Y/I$, the archived analysis gave
\begin{equation}
\Delta_Y=-4.15734\,\mu\Omega,
\qquad
u_{\Delta,Y}=1.04200\,\mu\Omega.
\end{equation}
The uncertainty term controlled the threshold in this comparison:
\begin{equation}
G_Y=3u_{\Delta,Y}=3.12601\,\mu\Omega.
\end{equation}
Thus $|\Delta_Y|>G_Y$, and the 3-mA field scan was withheld.
This was not the difference between the initial and final 1-mA
runs, and the failed screen did not uniquely distinguish heating,
non-ohmic response, contact changes, pickup or temporal drift.

\paragraph{Stationarity and detectability.}
The early baseline stationarity check compared the mean of the last
two blocks with that of the first two. For each component, the
acceptance threshold was
\begin{equation}
G_{a,\mathrm{stat}}=
\max\left[
3\sqrt{\frac{s_{a,\mathrm{early}}^{2}}{2}
       +\frac{s_{a,\mathrm{late}}^{2}}{2}},
\;0.01\sqrt{\bar r_X^2+\bar r_Y^2}
\right],
\end{equation}
where each $s$ is the sample standard deviation of the two block
means in the corresponding half. The screen passed when the
absolute late-minus-early difference did not exceed this threshold.

Detectability was assessed separately using
\begin{equation}
\mathrm{SNR}_{\mathrm{complex}}=
\frac{\sqrt{\bar r_X^2+\bar r_Y^2}}
     {\sqrt{u_X^2+u_Y^2}},
\end{equation}
with a recorded acceptance cutoff of 10. For the 2-K, zero-field
0.5-mA run, the component means were
$(-5.9593,60.4140)\,\mu\Omega$ and the block uncertainties were
$(6.9917,5.5346)\,\mu\Omega$. Therefore,
\begin{equation}
\mathrm{SNR}_{\mathrm{complex}}
=
\frac{\sqrt{(-5.9593)^2+(60.4140)^2}}
     {\sqrt{(6.9917)^2+(5.5346)^2}}
=6.8079.
\end{equation}
Although this run passed the descriptive half-to-half stationarity
screen, it failed the detectability criterion. Absence of a resolved
mean shift therefore did not qualify 0.5 mA as equivalent to the
reference setting. A separate screen for percentage magnetoresistance
required $|\bar r_X|\geq5s_{X,\mathrm{raw}}$, where
$s_{X,\mathrm{raw}}$ is the standard deviation of individual
current-normalized readings; this screen also failed for that run.

\paragraph{Correlation-aware timing qualification.}
The later 2-K, $+9$-T timing study compared two 61-reading traces
at each of the 1-s and 3-s lock-in time constants. Its protocol was
fixed before inspection of the high-field values. For each component
of the final 48 readings, mean-centered sample autocorrelations
$\rho(k)$ were calculated for lags $k=1,\ldots,10$. A correlation
scale and batch length were defined by
\begin{equation}
T_{\mathrm{corr}}=1+2\sum_{k=1}^{10}\max[0,\rho(k)],
\qquad
L_\tau=
\left\lceil
\max\left[4,\,2\max_{\mathrm{components,\ traces}}
T_{\mathrm{corr}}\right]
\right\rceil .
\end{equation}
This was a conservative operational heuristic, not a proven bound
on correlation. Irregular sampling, unresolved positive correlation
at lag 10 or failed tail-stationarity checks prevented qualification.

For an analysis window containing $N$ readings, all readings were
partitioned chronologically into $m=\lfloor N/L_\tau\rfloor$
nearly equal batches. With batch sizes $n_j$, means $b_j$ and
overall mean $\bar r$, the uncertainty estimate was
\begin{equation}
u_{\mathrm{within}}=
\max\left[
\sqrt{\frac{\sum_{j=1}^{m}n_j(b_j-\bar r)^2}{N(m-1)}},
\frac{s_{\mathrm{raw}}}{\sqrt{N}}
\right].
\end{equation}
At least four batches were required for a precision pass.
Stationarity additionally required the nominal 95\% Welch
$t$ interval for the late-minus-early half difference to lie
entirely within the selected margins of
$\pm20\,\mu\Omega$ for $X/I$ and $\pm25\,\mu\Omega$ for $Y/I$.
Each half required at least two batches.

Between-run variability was retained through a repeatability term
$s_{\mathrm{between}}$, calculated as the sample standard deviation
of the final-32-reading means of the two separate traces at each
time constant, provided those reference windows themselves qualified.
The total uncertainty for a subsequent single window was
\begin{equation}
u_{\mathrm{total},X}
=
\sqrt{u_{\mathrm{within},X}^{2}
+s_{\mathrm{between},X}^{2}}.
\end{equation}
The prospective precision target was
$u_{\mathrm{total},X}\leq5\,\mu\Omega$.
This was a selected standard-uncertainty target, not a 95\%
confidence-interval half-width or a threshold inferred from the
observed scatter. An unqualified repeatability term was treated as
unresolved, not set to zero.

The 30 analysis windows arose from two 1-s traces with three
candidate delays (25, 75 and 150 s), two 3-s traces with two
delays (75 and 150 s), and three window sizes (8, 16 and 32
readings): $2\times3\times3+2\times2\times3=30$.
Delay estimates used the settled-event timestamp minus the
programmed wait because the actual source-enable timestamp was
not recorded. All windows remained inconclusive under the full
acceptance rules; a small within-window uncertainty alone could
not qualify a faster setting. Independent follow-up qualification
was not completed, and no net time saving was established.

\subsection{Structured decision summaries}

The scientific record stores structured decision summaries rather than an
unrestricted narrative of private model reasoning. A summary identifies the
question being tested, cited evidence, competing explanations, proposed
discriminating measurement, applicable limits, validation result, observed
outcome and next action. This representation is both more reproducible and more
compact than a conversational transcript. It also permits a reviewer to
distinguish an observation from an interpretation and from a future proposal.

Supplementary Table~\ref{tab:decision_examples} gives a few examples from the TaP
measurements. A successful remeasurement is not treated as a
unique diagnosis: for example, increasing lock-in input headroom removed an
overload, but did not identify whether the original cause was an inappropriate
range, pickup, contact behavior or another transient. Similarly, a mean response
that appears unchanged is not promoted to an equivalence result when the stated
detectability criterion fails.

\begin{table}[p]
\centering
\caption{Examples of structured evidence-to-action decisions.
TaP qualification metrics and acceptance criteria are defined in
Section~\ref{sec:tap_qualification}.}
\label{tab:decision_examples}
\small
\begin{tabularx}{\linewidth}{@{}P{0.24\linewidth}Y@{}}
\toprule
Case & Evidence, guarded action and interpretation limit \\
\midrule

TaP overload, 300 K, 0 T &
\textbf{Evidence:} At $100\,\mu\mathrm{A}$ RMS and
\SI{177.77}{\hertz}, the SR860 overloaded with a
\SI{10}{\milli\volt} input range and
$10\,\mu\mathrm{V}$ sensitivity; no reading was accepted.
\textbf{Action:} Stop excitation, confirm source-off and
retest in a new validated plan with a \SI{1}{\volt} range
and \SI{50}{\milli\volt} sensitivity.
\textbf{Conclusion:} The retest produced 12/12 readable
values, establishing recovery but not the physical
overload cause \\

TaP current bracket, 100 K, 0 T &
\textbf{Evidence:} Within a 1--3--1 mA RMS bracket,
the 3-mA run and subsequent 1-mA return differed by
$4.157\,\mu\Omega$ in $Y/I$. Their combined descriptive
block uncertainty was $1.042\,\mu\Omega$, giving a
three-times-uncertainty threshold of
$3.126\,\mu\Omega$, which exceeded the alternative
1\% signal-scale floor. The observed difference
exceeded this threshold.
\textbf{Action:} Withhold the 3-mA field scan and return
to a longer-settled 1-mA plan.
\textbf{Conclusion:} The 1-mA scan yielded 180/180 valid
readings; heating, non-ohmic response, contact change,
pickup and drift were not separated, and 3-mA damage
was not established \\

TaP low-current bracket, 2 K, 0 T &
\textbf{Evidence:} In the 1--0.5--1 mA RMS comparison,
the 0.5-mA run gave
$\mathrm{SNR}_{\mathrm{complex}}
=\sqrt{\bar r_X^2+\bar r_Y^2}/\sqrt{u_X^2+u_Y^2}
=6.81$, where $\bar r_X$ and $\bar r_Y$ are the mean
current-normalized components and $u_X$ and $u_Y$
are their four-block standard uncertainties.
This fell below the selected detectability cutoff of 10.
\textbf{Action:} Do not promote 0.5 mA as an equivalent
high-field setting.
\textbf{Conclusion:} No mean shift was resolved, but
equivalence remained unproven because detectability
failed \\

TaP timing test, 2 K, +9 T &
\textbf{Evidence:} Four 61-reading traces compared
1-s and 3-s time constants. Two traces per setting,
three candidate delays for 1 s, two for 3 s and three
window sizes produced
$2\times3\times3+2\times2\times3=30$ analysis windows.
The selected precision target was
$u_{\mathrm{total},X}
=\sqrt{u_{\mathrm{within},X}^{2}
+s_{\mathrm{between},X}^{2}}
\leq5\,\mu\Omega$, subject to validity, correlation,
batch-count, stationarity and repeatability checks.
This target was a standard uncertainty, not a 95\%
interval half-width.
\textbf{Action:} Retain all traces and withhold a faster
production setting until all acceptance criteria pass.
\textbf{Conclusion:} All 30 windows were inconclusive;
unresolved repeatability was not treated as zero, and
neither a timing setting nor a net time saving was
qualified \\

\bottomrule
\end{tabularx}
\end{table}

The same record structure was used for a scientific inference in the
Co$_2$MnGa nonlinear-response campaign. At 20 K, reanalysis of the
0.5--1 mA amplitude series gave a descriptive power-law exponent
$p=2.034\pm0.056$, while comparison of the second-harmonic and DC-mixing
channels at matched output frequencies supported a common predominantly
field-odd quadratic response. Those observations motivated matched-field,
frequency and return controls, but did not distinguish an electronic nonlinear
Hall contribution from thermoelectric or other mixing mechanisms. The campaign
record therefore retains ``quadratic field-odd voltage'' as the supported result
and leaves the microscopic attribution unresolved.

\section{Experimental Examples of Agent-Guided Scientific Decisions}
\label{sec:experimental_decisions}

The following examples from the Co$_2$MnGa transport campaign
illustrate how QUANTA translates experimental evidence into
subsequent measurements. They demonstrate two complementary
capabilities: adapting a measurement plan based on acquired
data and investigating an unexpected response through
controlled diagnostic measurements. In both cases, scientific
planning is separated from validated instrument execution,
and physical interventions remain under human control.

\subsection{Adaptive measurement planning and run closure}
\label{sec:adaptive_measurement}

\begin{figure}[p]
    \centering
    \includegraphics[width=\linewidth]{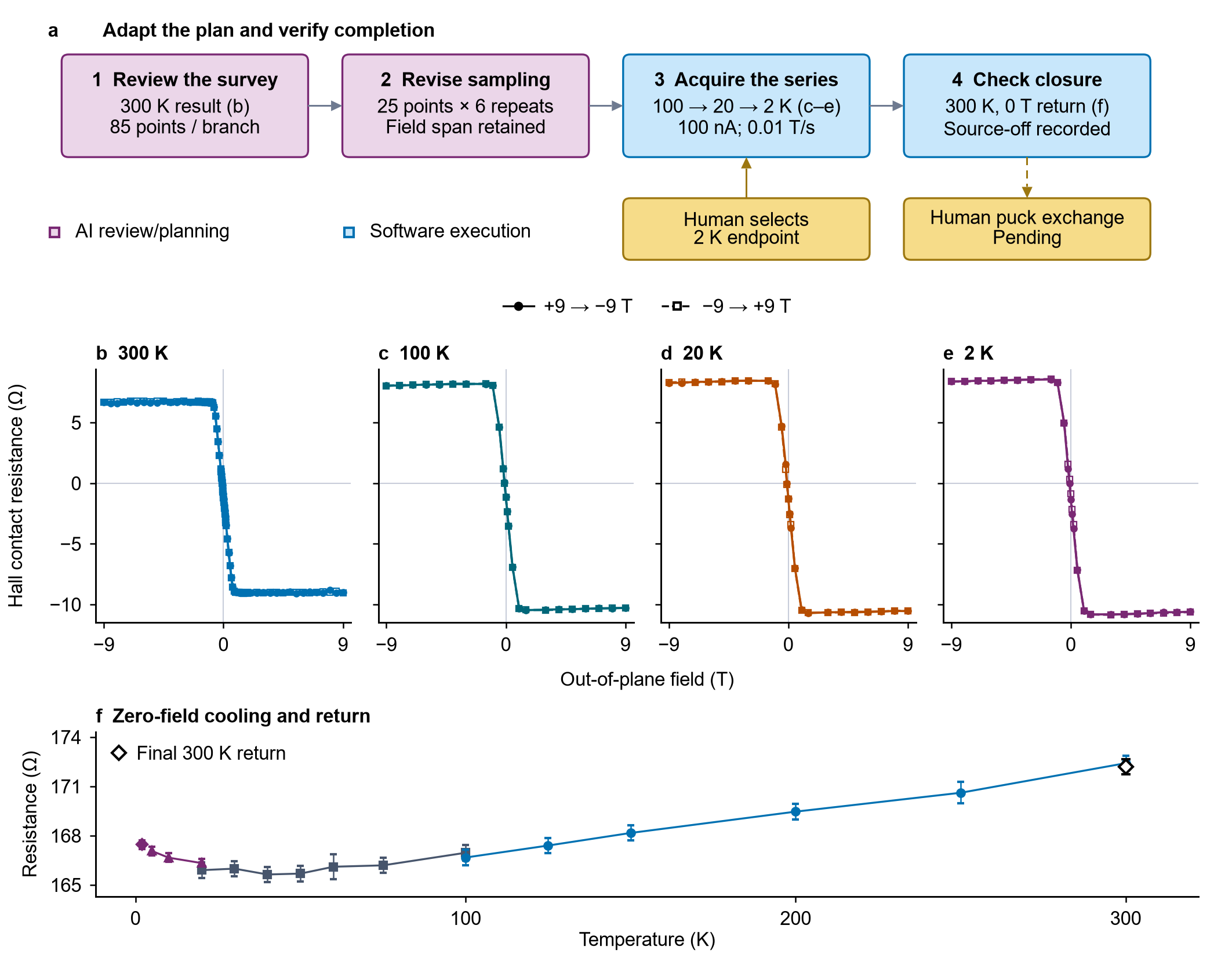}
    \caption{
    \textbf{Adaptive measurement planning and run closure
    in Co$_2$MnGa.}
    \textbf{(a)} Experimental decision sequence, from
    review of the initial 300 K survey to revised sampling,
    temperature-dependent acquisition and final-state checks.
    Purple denotes AI review and planning, blue denotes
    software execution, and yellow identifies human decisions
    or interventions.
    \textbf{(b)} Initial 300 K out-of-plane field sweep,
    acquired with 85 points per branch.
    \textbf{(c--e)} Subsequent measurements at 100, 20
    and 2 K using the revised sampling protocol of
    25 points and six repeats per point. Filled circles
    and open squares distinguish opposite field-sweep
    directions.
    \textbf{(f)} Zero-field resistance during the temperature
    sequence and final return to 300 K. The run concluded
    with recorded source-off and environmental return checks;
    the physical puck exchange remained pending.
    }
    \label{fig:si_adaptive_measurement}
\end{figure}

Supplementary Fig.~\ref{fig:si_adaptive_measurement}
illustrates an adaptive out-of-plane field measurement
sequence. After reviewing an initial 300 K survey with
85 points per field branch, the agent revised the sampling
strategy to 25 points with six repeats per point, preserving
the full field span while redistributing acquisition effort
toward repeated measurements. The subsequent series was
acquired at 100, 20 and 2 K with 100 nA excitation and a
field-sweep rate of 0.01 T\,s$^{-1}$. The 2 K endpoint
was selected by the researcher.

The resulting measurements show reproducible field-dependent
responses at the selected temperatures
(Supplementary Fig.~\ref{fig:si_adaptive_measurement}b--e).
The campaign concluded with a return to 300 K and zero
field, together with a recorded source-off state
(Supplementary Fig.~\ref{fig:si_adaptive_measurement}f).
These checks documented closure of the measurement run,
while the subsequent physical puck exchange remained a
pending human operation.

\subsection{Reference-assisted troubleshooting of an unexpected response}
\label{sec:reference_troubleshooting}

\begin{figure}[p]
    \centering
    \includegraphics[width=\linewidth]{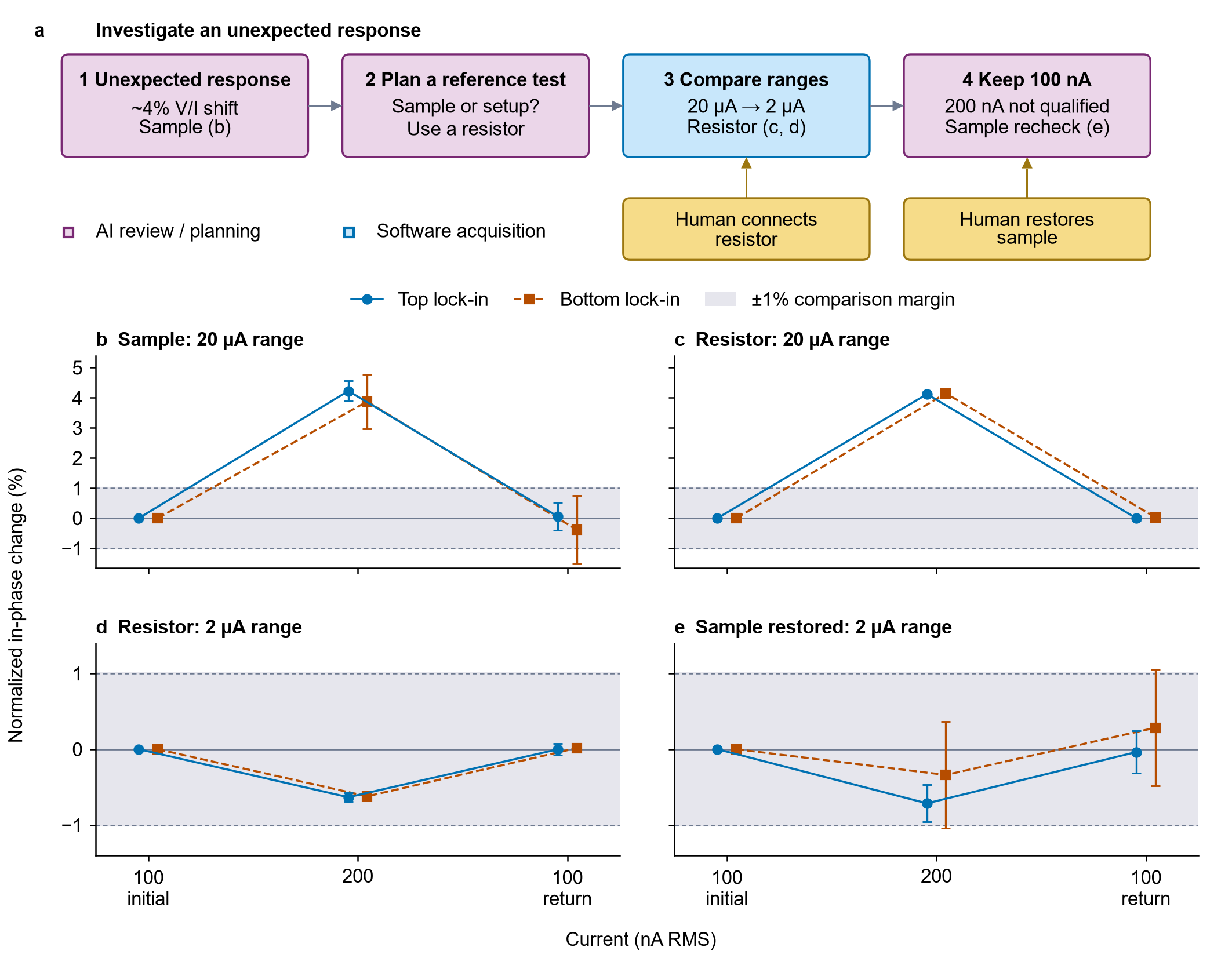}
    \caption{
    \textbf{Reference-assisted investigation of an unexpected
    transport response in Co$_2$MnGa.}
    \textbf{(a)} Diagnostic workflow initiated by an
    approximately 4\% change in the measured in-phase
    response. The agent proposed a reference-resistor
    comparison and evaluated the results, while a human
    operator performed the required physical connections.
    Purple indicates AI review and planning, blue indicates
    software acquisition, and yellow indicates human
    intervention.
    \textbf{(b)} Initial sample measurement using the
    20-$\mu$A current-source range.
    \textbf{(c)} Reference-resistor measurement with the
    same range, exhibiting a comparable response.
    \textbf{(d)} Reference-resistor measurement using the
    2-$\mu$A range, showing a substantially reduced
    variation.
    \textbf{(e)} Sample remeasurement after restoring the
    original wiring and using the 2-$\mu$A range.
    Blue circles and orange squares represent the two
    lock-in channels during 100--200--100 nA RMS
    excitation comparisons. The shaded region denotes
    the $\pm 1\%$ comparison margin. The diagnostic
    sequence supported retaining 100 nA but did not
    qualify 200 nA or uniquely establish the origin
    of the initial response.
    }
    \label{fig:si_diagnostic_test}
\end{figure}

A second example illustrates how the agent investigated an
unexpected change in electrical response rather than
immediately attributing it to the sample
(Supplementary Fig.~\ref{fig:si_diagnostic_test}).
An initial 100--200--100 nA current comparison produced
an approximately 4\% change in the normalized in-phase
voltage-to-current response. To distinguish possible
sample-related behavior from measurement-system effects,
the agent proposed repeating the comparison with a
reference resistor.

With the reference resistor connected by a human operator,
a comparable response was observed using the 20-$\mu$A
current-source range. Repeating the test with the narrower
2-$\mu$A range reduced the observed variation to within
the displayed $\pm 1\%$ comparison margin. After the
operator restored the sample, a further measurement
using the narrower range was performed. The resulting
evidence supported retaining 100 nA excitation, whereas
200 nA was not qualified for subsequent measurements.

This sequence illustrates evidence-driven troubleshooting
without requiring a unique diagnosis. The reference
measurements identified a dependence on the measurement
configuration but did not establish the microscopic or
instrumental origin of the initial shift. The agent
selected diagnostic steps and evaluated the resulting
evidence, while physical sample and resistor exchanges
required human intervention.

\section{Safety Rules and Action Envelopes}
\label{sec:safety}

\subsection{Layered safeguards}

Safety-critical behavior is implemented in deterministic software and explicit
campaign records rather than delegated to free-form text generation. Before
dispatch, the validator checks schema, units, point count, ordering, duration and
numerical ceilings. Dispatch additionally requires the configuration fingerprint
and current worker state to match the plan. During execution, the worker checks
temperature and field status, chamber status where available, source state,
compliance, lock-in reference and overload flags, finite readings and stop
latches. It attempts source-off on completion, pause, abort or fault and records
the resulting readback. Physical wiring, contact continuity, sample mounting and
the true sample current are not inferred from software telemetry.

Three boundaries are kept distinct. A \emph{hard limit} is imposed by the
deployed configuration, manufacturer constraint, user exclusion or verified
sample/contact bound. The \emph{authorized qualification envelope} is the range
within which a short, newly validated pilot may be attempted. The
\emph{demonstrated range} contains settings already supported by acceptable data.
A hardware rating or validator ceiling does not by itself move a setting into the
demonstrated range.

Supplementary Table~\ref{tab:envelopes} reports the numerical configurations
preserved with the selected campaigns. These values are campaign-specific and
must not be transferred to another specimen or wiring arrangement. They are
listed to make the action boundary auditable, not to imply that all extremes were
used simultaneously or are intrinsic damage thresholds.

\begin{table}[p]
\centering
\caption{Selected deployed action envelopes and experiment-specific gates.}
\label{tab:envelopes}
\small
\begin{tabularx}{\linewidth}{@{}P{0.25\linewidth}Y@{}}
\toprule
Campaign configuration & Recorded action envelope and interpretation \\
\midrule
TaP AC transport & \textbf{Environment:} 2--301 K and
$|B|\leq\SI{9}{\tesla}$; maximum rates \SI{2}{\kelvin\per\minute} and
\SI{0.01}{\tesla\per\second}. \textbf{Electrical:}
$I_{\mathrm{peak}}\leq\SI{4.3}{\milli\ampere}$, compliance
$\leq\SI{0.1}{\volt}$ and configured estimated power
$\leq10\,\mu\mathrm{W}$. \textbf{Run bound:} 3,000 points, 50,000 readings and
24 h. The current ceiling was a conservative qualification boundary, not a
demonstrated sample-safe limit \\
Co$_2$MnGa out-of-plane Hall & \textbf{Environment:} 2--300 K and
$|B|\leq\SI{9}{\tesla}$; maximum rates \SI{3}{\kelvin\per\minute} and
\SI{0.02}{\tesla\per\second}. \textbf{Electrical:}
$I_{\mathrm{peak}}\leq0.30\,\mu\mathrm{A}$, compliance
$\leq\SI{0.1}{\volt}$ and configured estimated power
$\leq\SI{10}{\nano\watt}$. \textbf{Run bound:} 10,000 points, 100,000
readings and 24 h. Both voltage channels were read sequentially; a transverse
channel label did not by itself establish Hall polarity \\
Co$_2$MnGa in-plane nonlinear campaign & \textbf{Environment:} 2--300 K and
$|B|\leq\SI{1}{\tesla}$; maximum rates \SI{4}{\kelvin\per\minute} and
\SI{0.01}{\tesla\per\second}. \textbf{Electrical:}
$I_{\mathrm{peak}}\leq\SI{1.5}{\milli\ampere}$, compliance
$\leq\SI{5}{\volt}$ and port-power screening $\leq\SI{15}{\milli\watt}$;
commissioned transport used at most \SI{1}{\milli\ampere} RMS with a
\SI{180}{\second} source-on timer. \textbf{Run bound:} 10,000 points, 100,000
readings and 6 h. Port-power screening was not a measurement of sample or
contact dissipation \\
\bottomrule
\end{tabularx}
\end{table}

\subsection{Fault classification and response}

The system separates electrical protection, acquisition validity and scientific
precision. A valid but noisy reading may support a limited observation; an
overloaded or unlocked reading cannot. Supplementary
Table~\ref{tab:fault_response} summarizes the main classes considered in the
troubleshooting procedure. Only changes within the recorded envelope can be
tested automatically. Any branch requiring rewiring, sample exchange, contact
repair or another unverifiable physical action is held for a person.

\begin{table}[htbp]
\centering
\caption{Fault and data-quality classes used in the supervisory workflow.}
\label{tab:fault_response}
\small
\begin{tabularx}{\linewidth}{@{}P{0.27\linewidth}Y@{}}
\toprule
Observation class & Response and remaining limitation \\
\midrule
Stale deployment, owner mismatch or unverified wiring map & Hold dispatch;
reconcile worker, configuration and physical record. \textbf{Unresolved:}
actual lead continuity, grounding and sample identity require independent evidence \\
Malformed or out-of-envelope plan & Reject before queueing and return the failed
field or numerical bound. Scientific desirability cannot override a hard limit \\
Source compliance, overtemperature or inconsistent source state & Stop the
affected run, preserve flags and attempt verified source-off. \textbf{Unresolved:}
programmed current is not an independent measurement of sample current \\
SR860 input overload or reference loss during excitation & Reject affected
readings, stop when required and use a new validated diagnostic plan.
\textbf{Unresolved:} a range change can restore acquisition without identifying
the physical cause \\
Chamber fault, non-finite or persistently unknown $T/B$ state & Pause or fault,
attempt source-off and retain the last PPMS targets for explicit review.
\textbf{Unresolved:} chamber mode is not a pressure measurement; electrical
shutdown does not zero the magnet \\
Communication or worker fault & Preserve durable run state, block blind restart
and reconcile ownership before a new run. \textbf{Unresolved:} process or bus
failure can prevent a guaranteed shutdown; this is not an emergency-stop system \\
Operator pause, abort or stop & Honor the latch at a polling boundary, block
future stage dispatch and verify the physical outcome. \textbf{Unresolved:} a
stop acknowledgment alone is not proof of the final apparatus state \\
Valid but noisy, drifting or under-resolved data & Retain observations with
uncertainty; choose a bounded current, frequency, settling or return control only
if it can change the conclusion. \textbf{Unresolved:} longer averaging cannot
repair overload, reference loss or systematic drift \\
Required physical intervention & Keep excitation off and request a labeled
operator action, then requalify the affected path. Software cannot perform or
verify rewiring, contact repair or sample exchange \\
\bottomrule
\end{tabularx}
\end{table}

\section{Dashboard and Measurement Monitoring}
\label{sec:dashboard}

\begin{figure}[p]
    \centering
    \includegraphics[width=0.88\linewidth]{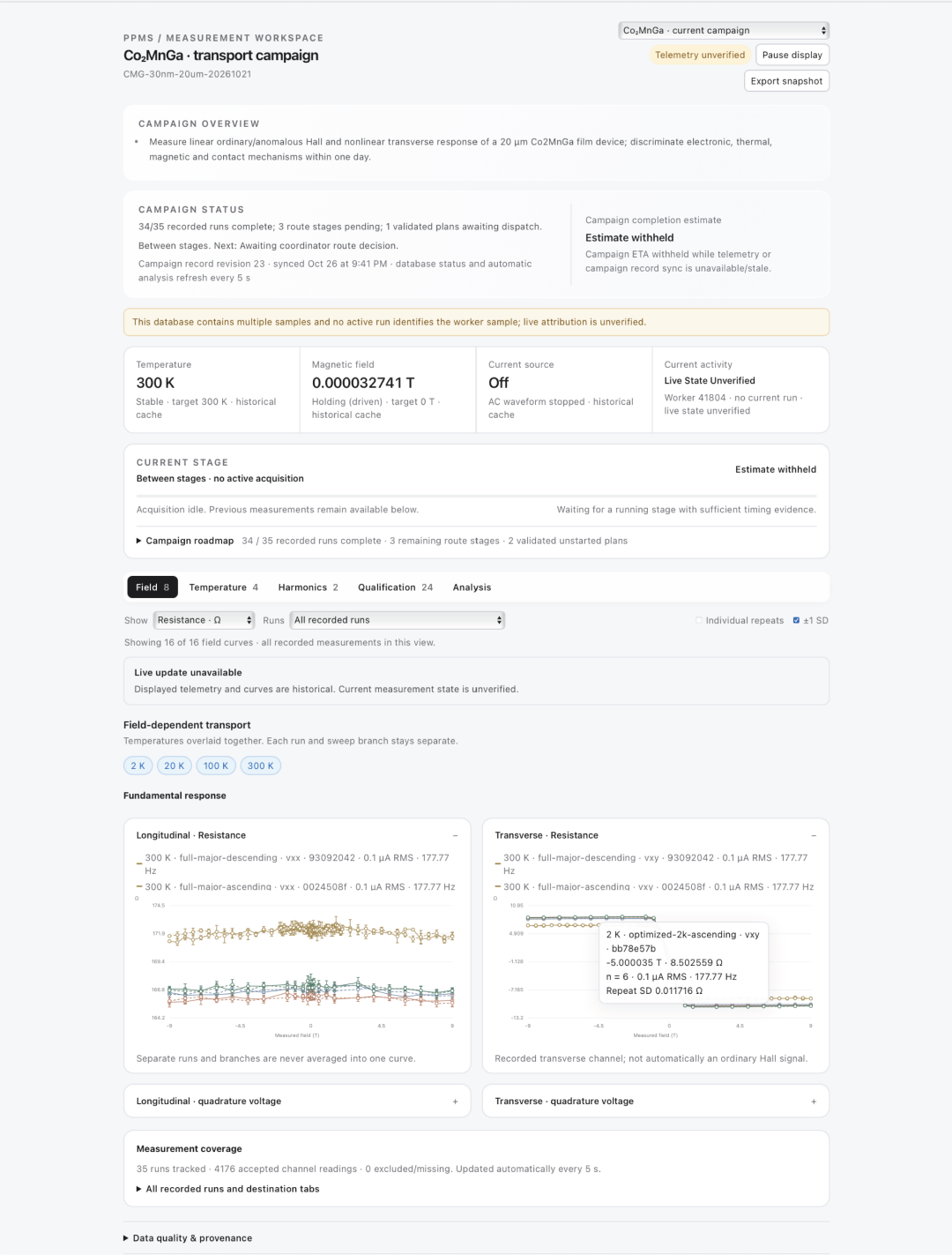}
    \caption{
        \textbf{Read-only campaign dashboard.} Example Co$_2$MnGa transport
        workspace showing the campaign objective, recorded-run progress, roadmap,
        cached temperature/field/source state, measurement tabs, individual field
        branches and provenance controls. The visible ``telemetry unverified''
        and ``estimate withheld'' messages illustrate the freshness policy: the
        interface does not present historical cache values as a verified live
        instrument state. The dashboard has no instrument-control path.}
    \label{fig:si_gui}
\end{figure}

The browser dashboard is a read-only projection of committed campaign records
(Supplementary Fig.~\ref{fig:si_gui}). It displays the campaign objective,
progress, current stage, cached temperature, field and source state, pending
validated plans, data-quality annotations and versioned analysis products. Run
branches and individual repeats remain separable rather than being silently
averaged into a single curve. Freshness warnings are part of the interface: when
the database cannot be attributed to a verified live worker, the dashboard marks
telemetry as historical or unverified and withholds an estimated completion time.
Registration and display metadata select what is shown, but cannot select an
instrument endpoint or authorize dispatch.

\section{Retrieval Corpus, Coverage and Leakage Controls}
\label{sec:retrieval}

\begin{figure}[htbp]
    \centering
    \includegraphics[width=\linewidth]{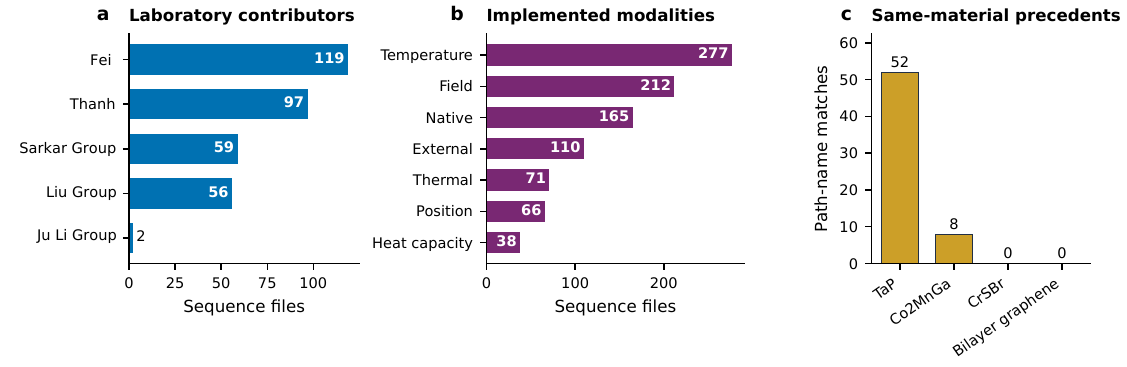}
    \caption{
        \textbf{Composition of the laboratory sequence corpus.}
        \textbf{(a)} Number of \texttt{.seq} files by contributor directory.
        \textbf{(b)} Multi-label modality counts based on active command tokens;
        one file may contribute to several bars. \textbf{(c)} Direct
        case-insensitive material-name matches in sequence paths for the four
        manuscript examples. TaP and Co$_2$MnGa therefore require explicit
        same-material exclusion in a held-out evaluation. }
    \label{fig:sequence_corpus}
\end{figure}

The archived laboratory-sequence corpus contains 333 native \texttt{.seq} files
from five contributor directories (Supplementary
Fig.~\ref{fig:sequence_corpus}a). Active command tokens show broad procedural
coverage: 277 files contain temperature control, 212 field control, 165 native
resistance or transport commands, 110 external electrical-transport commands,
71 thermal-transport commands, 66 position or rotation commands and 38 heat-
capacity commands (Supplementary Fig.~\ref{fig:sequence_corpus}b). These are
multi-label counts because one sequence can combine several modules and control
axes. Disabled lines beginning with \texttt{!} were excluded from the command
classification.

The corpus is useful as procedural evidence---for example, for identifying
settling patterns, nested field and temperature loops and return branches---but
it is
not an unbiased evaluation set. Case-insensitive path-name matching finds 52 TaP
sequences and 8 Co$_2$MnGa sequences 
(Supplementary Fig.~\ref{fig:sequence_corpus}c). The TaP and Co$_2$MnGa measurements
therefore have direct same-material precedents and cannot be presented as
material-held-out tests of generalization. Absence of a material name also does
not prove absence of indirect leakage: a Hall or temperature-sweep protocol from
another material can still encode a useful route.

For a prospective generalization benchmark, retrieval indices and evaluation
records should be split before the run at the specimen, material-family,
contributor and time levels. A strict material-held-out split would exclude all
TaP-named files when evaluating TaP and all Co$_2$MnGa-named files when evaluating
Co$_2$MnGa. A second, protocol-held-out split should remove near-duplicate Hall,
MR or temperature sequences even when their material names differ. The retrieval
identifiers and hashes returned to the agent should be logged, and performance
reported both with and without same-material precedents. The present archive does
not contain a completed prespecified held-out evaluation; consequently, the
sequence analysis documents coverage and leakage risk rather than establishing a
generalization score.

\section{Reproducibility, Provenance and Scope}
\label{sec:provenance}

Both the TaP and Co$_2$MnGa campaigns used OpenAI Astra Ultra for the Codex coordinator and specialist agents. The coordinator organized the campaigns and delegated tasks, while specialist agents supported experimental planning, troubleshooting and data interpretation.

\subsection{Evidence archive and interpretive limits}
\label{sec:evidence_archive}

The archived evidence bundle preserves the control software, deployment
configurations, tests, laboratory sequences, campaign records, structured plans,
SQLite run databases, raw JSON/JSONL exports, analysis scripts, dashboard
snapshots and final reports. Figure-generating analyses for Supplementary
Figs.~\ref{fig:sequence_corpus} are kept outside
the source-data tree and read the archive without modifying campaign records.
The derived summary records the source-archive hash, corpus classification rules,
material-name matches and the exact uniform-grid
comparison.

The record has several deliberate limits. Instrument readbacks are sequential,
not hardware-synchronous. A programmed current is not an independent sample-
current measurement. PPMS stability does not by itself establish sample
equilibrium, and chamber mode does not establish pressure. Source-off attempts do
not constitute an emergency stop and do not cancel retained temperature or field
targets. The TaP geometry record does not support an absolute resistivity without
additional dimensions and contact calibration. Finally,
the present corpus audit identifies same-material leakage but does not substitute
for a prospective held-out benchmark. These boundaries are retained so that the
technical description does not imply capabilities or scientific conclusions that
the archived measurements do not support.

\subsection{Agent activity and computational usage}
\label{sec:agent_activity}

\begin{figure}[p]
    \centering
    \includegraphics[
        width=\linewidth,
        height=0.85\textheight,
        keepaspectratio
    ]{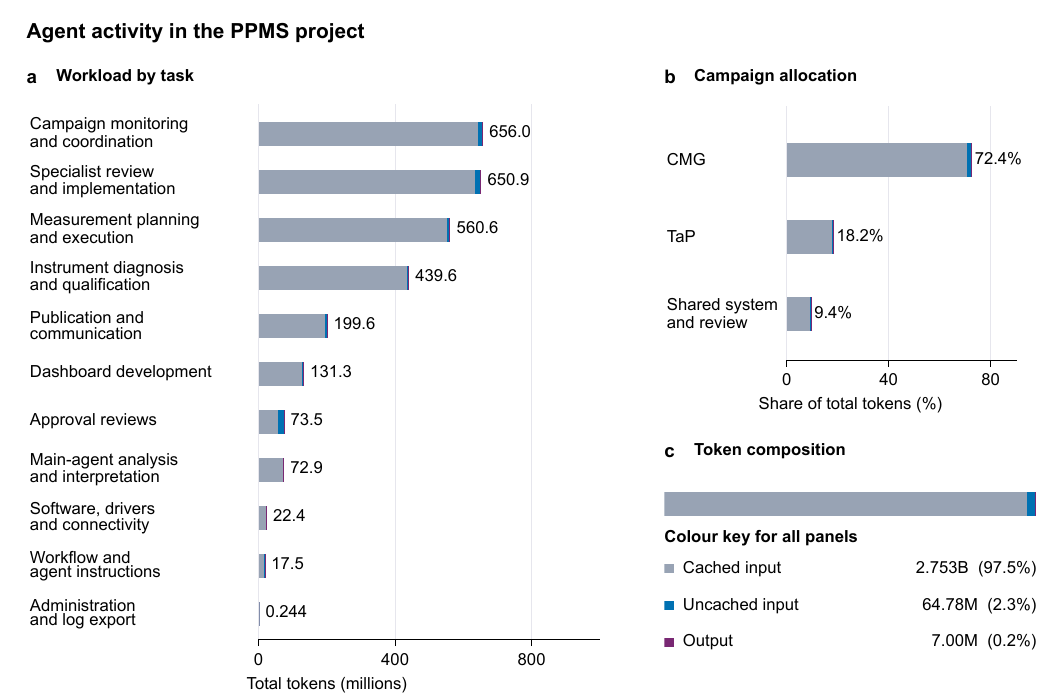}
    \caption{
    \textbf{Agent activity and token usage in the PPMS project.}
    \textbf{(a)} Cumulative token consumption by task category,
    including campaign monitoring and coordination, specialist
    review and implementation, measurement planning and execution,
    instrument diagnosis and qualification, and supporting
    activities. Values are reported in millions of tokens.
    \textbf{(b)} Allocation of total token usage across the
    Co$_2$MnGa (CMG) and TaP,
    together with shared system and review activities.
    \textbf{(c)} Aggregate token composition, comprising
    2.753 billion cached-input tokens (97.5\%),
    64.78 million uncached-input tokens (2.3\%) and
    7.00 million output tokens (0.2\%). Colors distinguish
    cached input, uncached input and output across all panels.
    The statistics represent cumulative recorded model usage,
    including development and supporting activities, rather
    than independent reasoning steps or direct measures of
    scientific performance.
    }
    \label{fig:si_token_usage}
\end{figure}

Supplementary Fig.~\ref{fig:si_token_usage} summarizes the recorded
agent activity across the PPMS project, including experimental campaigns,
software development, instrument qualification and supporting tasks.
The largest token expenditures were associated with campaign monitoring
and coordination, specialist review and implementation, measurement
planning and execution, and instrument diagnosis and qualification.
Among the experimental campaigns, Co$_2$MnGa accounted for the largest
share (72.4\%), followed by TaP (18.2\%).

The recorded usage totaled approximately 2.75 billion tokens, of which
97.5\% were cached input, 2.3\% uncached input and 0.2\% output.
The dominance of cached input reflects substantial reuse of context
across agent interactions. These cumulative counts include development
and supporting activities and should not be interpreted as independent
scientific reasoning steps, unique input content, or direct measures of
experimental efficiency. They instead provide a retrospective account
of the scale and distribution of model usage.

